\documentclass[a4paper, 12pt]{article}
\usepackage[utf8]{inputenc}
\usepackage{amsmath, amssymb}
\usepackage{graphicx}
\usepackage{caption}
\usepackage{booktabs}
\usepackage{float}
\usepackage{geometry}
\usepackage{lipsum} 
\usepackage{hyperref}
\usepackage{lscape}
\usepackage{cleveref} 
\usepackage{algorithm}      
\usepackage{algpseudocode}  

\usepackage{pdflscape}
\usepackage{longtable}
\usepackage{array}

\hypersetup{
    colorlinks=true,
    linkcolor=blue,
    filecolor=magenta,      
    urlcolor=cyan,
    citecolor=blue,         
    pdftitle={Research Paper},
    pdfpagemode=FullScreen,
}

\title{Forecasting in small open emerging economies: Evidence from Thailand}
\author{
    Paponpat Taveeapiradeecharoen\thanks{PhD, paponpat.tav@mfu.ac.th} \and Nattapol Aunsri\thanks{computer engineering, nattapol.aun@mfu.ac.th}
}
\date{\today}

\usepackage[authoryear,longnamesfirst]{natbib}
\begin{document}

\date{\today}
\maketitle

\begin{abstract}
Forecasting inflation in small open economies is difficult because limited time series and strong external exposures create an imbalance between few observations and many potential predictors. We study this challenge using Thailand as a representative case, combining more than 450 domestic and international indicators. We evaluate modern Bayesian shrinkage and factor models, including Horseshoe regressions, factor-augmented autoregressions, factor-augmented VARs, dynamic factor models, and Bayesian additive regression trees.

Our results show that factor models dominate at short horizons, when global shocks and exchange rate movements drive inflation, while shrinkage-based regressions perform best at longer horizons. These models not only improve point and density forecasts but also enhance tail-risk performance at the one-year horizon. 

Shrinkage diagnostics, on the other hand, additionally reveal that Google Trends variables, especially those related to food essential goods and housing costs, progressively rotate into predictive importance as the horizon lengthens. This underscores their role as forward-looking indicators of household inflation expectations in small open economies.

\end{abstract}

\section{Introduction}
Inflation forecasting in small open economies presents persistent challenges. These economies face limited time series length and high exposure to external shocks, while at the same time policymakers must monitor a wide range of predictors such as domestic activity, labor market slack, commodity prices, exchange rates, and global financial conditions. Standard low-dimensional models often struggle in such environments because they cannot flexibly balance large sets of predictors against relatively few observations. This motivates the implementation of high-dimensional Bayesian shrinkage methods, such as the Horseshoe prior, and factor-based approaches. The Horseshoe prior in particular is designed to handle sparse signals in high-dimensional data, allowing the model to aggressively shrink irrelevant predictors while preserving large coefficients. This makes it especially well-suited for small open economies, where the number of potential predictors can far exceed the available sample size.

This study develops a unified framework that compares state-of-the-art Bayesian and factor models in forecasting inflation for small open economies, using Thailand as a representative case. Thailand is particularly suitable because it is an emerging market highly integrated into global trade and commodity networks, yet it also experiences episodes of volatility from domestic shocks. Lessons from this setting can generalize to other open economies in Southeast Asia and beyond, where policy authorities must contend with similar forecasting difficulties.

Our contributions lie for policymakers who are related to forecasting. First we construct a reproducible pipeline that benchmarks Horseshoe regression, Factor-Augmented AR, Factor-Augmented VAR, Dynamic Factor Models, and Bayesian Additive Regression Trees under a common rolling evaluation. Secondly we introduce a shrinkage diagnostic based on posterior shrinkage ratios from the Horseshoe prior, which allows us to track changing drivers of inflation. Next we also provide a comprehensive density forecast evaluation using CRPS, log scores, and quantile-weighted scores that highlight model performance in left, right-tail and simultaneous tails outcomes. Finally we offer practical guidance on when direct versus iterated factor forecasts are most reliable in data-rich environments. While our empirical evidence focuses on Thailand, the framework is designed to speak to forecasting strategies in small open economies more generally.

Our roadmap for this work can be summarised as followed: First is \cref{sec:data} describes data and transformations. \Cref{sec:models} details models. \Cref{sec:design} explains the rolling design and metrics. \Cref{sec:results} reports results and Diebold-Mariano (DM) tests. \Cref{sec:drivers} analyzes drivers via shrinkage factor. \Cref{sec:conclude} concludes.

\section{Data and Transformations}
\label{sec:data}
The forecasting dataset combines an extensive collection of Thai and international macroeconomic indicators, financial market variables, and measures of household expectations. This mix of domestic indicators with global drivers reflects the reality of small open economies, where local inflation dynamics cannot be separated from international trade, commodity prices, and global financial shocks.

The primary source is the Bank of Thailand (BoT), which maintains its own statistics and consolidates data from government agencies including the Ministry of Commerce, the Department of Lands, the Revenue Department, the National Statistical Office, and the Social Security Office. To account for global drivers of inflation, we supplement the Thai series with commodity prices, financial market indicators, and U.S. macroeconomic aggregates drawn from FRED, IMF primary commodity statistics, and Yahoo Finance. Such augmentation is particularly important for small open economies, where external conditions and global price shocks can transmit quickly into domestic inflation.

We also include Google Trends search volumes\footnote{Google trend typically publish these volumn by rescaling them into 0-100, so search volumes here not strictly means the number of total search.} for terms like "egg price", "rent price", and "boxed meal price". These variables do not measure formal inflation expectations like survey-based or market data. Instead, they capture the attention and concern of consumers about common price items. In many small open economies, including ours, survey data on inflation expectations are quite often limited or unavailable. So search behavior serves as a valuable, real-time signal of perceived cost-of-living pressures. That said, these indicators reflect behavioral attention, not literal forecasts of future inflation. Still, they are useful real-time perceived inflation stress and supplement our panel of domestic and global variables effectively, see for instance \citep*{matheson2010analysis,castelnuovo2017google}.

All series are sampled at monthly frequency. The sample period begins in the late 1990s, although the precise start date varies across series depending on availability. Missing values are minimal, and in cases where they occur at the beginning or end of a series, we retain the series after transformation to preserve information. The final balanced panel contains almost over 500 predictors spanning domestic activity, consumption, investment, trade, labor markets, credit and property markets, exchange rates, external prices, and global financial conditions. This breadth of variables mirrors the information environment faced by many small open economies, which must process both limited domestic data and extensive global signals. The ultra-high-dimensional setting provides an ideal laboratory for Bayesian shrinkage and factor-based methods. This high-dimensional Bayesian shrinkage is particularly suited to small open economies that have limited observations but many predictors, see for instances \citep*{huber2019adaptive,Nookhwun2023}. A complete list of variables, their sources, and transformation codes is provided in the online Supplementary Catalog (see Online Supplementary Material).  

Transformations follow the \citet{mccracken2016fred} benchmark protocol of \textit{FRED-MD}, which is widely used in empirical macroeconomic forecasting. Each raw series $x_t$ is transformed to achieve covariance stationarity while maintaining economic interpretability. Price and quantity indexes are expressed in log first differences ($\Delta \log x_t$), approximating monthly growth rates. Levels are retained for interest rates, spreads, and bounded indexes that are stationary by construction. Simple first differences are applied to ratios and flow series that are not meaningful in logs, as well as to survey balances that contain nonpositive values. When unit root evidence remains after first differencing, higher-order differencing is applied, though such cases are rare.  

Transformation choices are guided by both statistical tests and economic rationale. For example, exchange rates and equity prices are entered in log differences, while survey-based sentiment indexes are left in levels since they are already bounded. This approach ensures comparability across predictors and improves interpretability of posterior shrinkage patterns in the Bayesian models. For transparency, the Supplementary Catalog not only reports the assigned transformation code for each variable but also records the rationale underlying the decision rule.

\section{Models}\label{sec:models}
Our selection of models is guided by three complementary principles for forecasting in data-rich environments that characterize small open economies, such as Thailand: shrinkage, factors, and flexibility. The monthly panel contains on the order of five hundred predictors, many of which move together because they reflect common domestic and global forces. A single class of models is unlikely to dominate across all horizons \(h\in\{1,3,6,12\}\), so we evaluate archetypes that operationalize distinct ways to extract signal while respecting publication lags and avoiding look-ahead.

The first principle is high-dimensional shrinkage. The horseshoe regression provides a direct, horizon-specific map from \(x_{t-L}\) to \(y_{t+h}\) with global-local regularization that can both suppress noise and retain a few strong signals. It is designed for the \(p\gg n\) regime and yields full predictive densities together with interpretable shrinkage diagnostics \((1-\kappa)\) that we exploit to trace time-varying drivers. This makes it especially suitable for small open economies where policymakers must process hundreds of domestic and international predictors despite having relatively short macroeconomic time series, as recently pointed out by \cite{huber2019adaptive}. As a benchmark and for the sake of relative skill, we also keep a transparent AR baseline estimated under a flat prior.

The second principle is dimension reduction through factors. When many predictors share common variation, principal-components factors offer a parsimonious representation. In small open economies, such factor structures capture the influence of global commodity prices, exchange rates, and regional demand that often move together and dominate domestic inflation dynamics, among others \citep*{stock2002macroeconomic,crucini2008persistence}. Another evidence from specifically small open economies \citep{aastveit2016world} shows that global and regional shocks significantly shape cyclical dynamics. With all these in mind, we therefore include a factor-augmented regression (FA-AR) that projects \(y_{t+h}\) directly on estimated factors and lags of \(y_t\), and a factor-augmented VAR (FAVAR) that models the joint dynamics of factors and inflation and produces multi-step forecasts by iteration. The direct specification allows horizon-by-horizon shrinkage of the mapping and typically excels at short horizons; the iterated specification lets factor dynamics accumulate and can be advantageous at medium and longer horizons. A dynamic factor model (DFM) complements these by placing the factor structure in a state-space form with explicit measurement noise and a transition for the latent factors, delivering iterated forecasts via the Kalman filter. In practice, FA-AR, FAVAR, and DFM speak to the same economic idea—that a small number of latent forces summarize broad comovement—but they differ in how that idea is operationalized for forecasting.

The third principle is functional flexibility. Relationships between inflation and predictors can be nonlinear or interact in ways that linear shrinkage and static factors may miss, especially around commodity or exchange-rate shocks. We therefore include Bayesian Additive Regression Trees (BART), a machine-learning specification that approximates unknown nonlinear functions by a sum of shallow trees with Bayesian regularization. BART produces full predictive distributions and provides a useful counterpoint to linear shrinkage and factor models in the same evaluation design.

We compare both direct and iterated forecasting because they address different bias-variance trade-offs. Direct models estimate the \(h\)-step mapping explicitly and can reduce accumulation of dynamic misspecification at short horizons, but they do not exploit cross-equation restrictions. Iterated models borrow strength from an estimated law of motion for the state, which can help as \(h\) grows but may compound model error. Our rolling, expanding-window design puts all six specifications on the same footing: identical transformations and standardization computed within each training window, the same publication delay \(L\), the same forecast origins and horizons, and evaluation by both point and density criteria. This unified setup lets us isolate what each principle-shrinkage, factors, and flexibility-buys for Thai inflation forecasting, and how their relative merits shift across horizons.

\subsection{Autoregressive baselines}
As a transparent benchmark we use horizon-specific direct autoregressions on the transformed target with (uninformative prior), letting those likelihood dominate the conditional posterior distribution. For each horizon \(h\) and origin \(t\), the direct AR(\(p\)) writes
\begin{equation}
\label{eq:mainREG}
y_{t+h} \;=\; \alpha_h \;+\; \sum_{i=1}^{p} \phi_{h,i}\, y_{t+1-i} \;+\; \varepsilon_{t+h},
\qquad \varepsilon_{t+h}\sim\mathcal N(0,\sigma_h^2),
\end{equation}
so the regressors are \(x_t=[y_t,\,y_{t-1},\ldots,y_{t-p+1}]'\). This "direct" mapping is estimated recursively with expanding windows and respects the information set at each origin. In practice we report AR(2) as the canonical baseline. Using an AR benchmark in inflation forecasting is standard and facilitates comparability with the literature; see, e.g., \citep{stock1999forecasting,stock2008phillips} and \citep{faust2013forecasting}.

Estimation adopts a flat Bayesian prior as followed:
\[
p(\beta,\sigma^2)\ \propto\ \frac{1}{\sigma^2}, 
\qquad \beta=\big(\alpha_h,\phi_{h,1},\ldots,\phi_{h,p}\big)' ,
\]

i.e., flat in \(\beta\) and Jeffreys in \(\sigma^2\). With \(X\) the \(n\times(p{+}1)\) design matrix built from \([1,y_t,\ldots,y_{t-p+1}]\) over the training sample and \(y\) the stacked \(y_{t+h}\), the posterior is conjugate and coincides with OLS in mean \citep{zellner1996introduction,koop2003bayesian}:
\begin{align}
\beta \mid \sigma^2,y,X &\sim \mathcal N\!\big(\hat\beta_{\text{OLS}},\, \sigma^2 (X'X)^{-1}\big),\\
\sigma^2 \mid y,X &\sim \text{Inv-Gamma}\!\Big(\tfrac{n-k}{2},\, \tfrac{\text{SSE}}{2}\Big),
\end{align}
where \(k=p{+}1\), \(\hat\beta_{\text{OLS}}=(X'X)^{-1}X'y\), and \(\text{SSE}=(y-X\hat\beta_{\text{OLS}})'(y-X\hat\beta_{\text{OLS}})\). The one-step-ahead predictive for a new regressor \(x_{\mathrm{oos}}\) is Student-\(t\):
\begin{equation}
y_{\mathrm{oos}} \mid y,X \ \sim\ t_{\,n-k}\!\Big( x_{\mathrm{oos}}'\hat\beta_{\text{OLS}},\,
\hat\sigma^2 \big(1 + x_{\mathrm{oos}}'(X'X)^{-1}x_{\mathrm{oos}}\big) \Big),
\qquad \hat\sigma^2=\text{SSE}/(n-k),
\end{equation}
This baseline is attractive because it is fully explicit, numerically stable in small \(n\), and widely used as a yardstick in inflation forecasts \citep{stock1999forecasting,stock2008phillips,faust2013forecasting}. It also provides a neutral reference for relative skill scores reported later.

\subsection{Ultra-high-dimensional Bayesian HS (Direct)} 
This is probably our main model to be competitive with plenty of previous successful models to handle the high-dimensional predictors factor-augmented regression, and VAR (FA-AR, FAVAR), so forth and so on which will be described shortly after this sub-section. We emphasize that this ultra-high-dimensional setup is not unique to Thailand but generalizes to many small open economies, where the available number of observations is dwarfed by the set of potentially relevant predictors. Here we introduce for convenience. For each horizon $h$, we estimate similarly as described in \cref{eq:mainREG} but with large amount of predictors rather than simply just inflation's lag(s).
\begin{align*}
y_{t+h} = x_{t}'\beta + \varepsilon_{t+h},\qquad \varepsilon_{t+h}\sim \mathcal N(0,\sigma^2),
\end{align*}
where $x_{t}$ contains all $p$ predictors and denote $n$ as total number of observations. Because $k$ can exceed $n$ by an order of magnitude, potentially contain all source of inflation movement and thus hopefully to improve out rolling expanding windows out-of-sample forecast. 

Like we have described above that our predictors are in the state of ultra-high-dimensional relative to its number of observations we need sampling method to avoid near singular matrix after the inverse of the term $X'X$ during the regression coefficient sampling. To avoid such problem we do implement the fast sampling method pioneered by \cite{bhattacharya2016fast}. Computationally, the key step avoids inverting $X'X$ directly by sampling a Gaussian auxiliary vector, solving an $n \times n$ linear system \( Aw=(y-v)\), then \(A=X (\sigma^2\tau^2\Lambda^2)X'+\sigma^2I_n\) and then recovering \(\beta = u + \sigma^2\tau^2\Lambda^2X'w\). As a result the draws for $\sigma^2$, $\lambda_j^2$, and $\tau^2$ follow from conjugate full conditionals. A compact summary of the sampler is provided in \cref{alg:bhatt-fast-beta}. 
The fast sampler of \citet{bhattacharya2016fast} is essentially \emph{prior-agnostic} for the coefficient step: once the prior implies a diagonal covariance $D=\sigma^2\tau^2\Lambda^2$ (or, more generally, any diagonal scale matrix), the $\beta$–update in \cref{alg:bhatt-fast-beta} goes through unchanged and requires solving only an $n\times n$ system. Global–local priors then differ only in their \emph{scale} updates. To avoid over-fitting we impose the horseshoe prior \citep{carvalho2010horseshoe}. Such prior is popular among econometric field research and successfully prove to handle over-fitting quite well and one worth advantage worth noting is that they are predetermined hyper-parameter free, see for examples, \cite{cross2020macroeconomic,gefang2022forecasting,huber2023nowcasting}. 
\begin{align}
\label{eq:hspriors}
    \beta_j \mid \lambda_j, \tau, \sigma^2 \sim \mathcal{N}\!\left(0, \, \sigma^2 \tau^2 \lambda_j^2\right),
\end{align}
In particular, replacing the horseshoe with alternatives such as the normal–gamma, Dirichlet–Laplace, or $R^2$–D$^2$ priors amounts to swapping the conditional draws for the local and global scales, while keeping the same fast $\beta$–draw. This modularity makes the approach well suited to $k\gg n$ panels: numerical stability is improved (no inversion of $X'X$), memory demands are modest, and the sampler is easily adapted across shrinkage families. In summary we adopt the horseshoe prior \citep{carvalho2010horseshoe} with Makalic-Schmidt updates for the local $\lambda_j^2$ and global $\tau^2$ scales, see \citep{makalic2015simple}. The resulting one sweep Gibbs iteration (ultra-high-dimensional and fast $\beta$, then $\sigma^2$, then local and global scales) is summarized in \cref{alg:hs-gibbs}.



\begin{algorithm}[H]
  \caption{Fast $\beta$ draw for high-dimensional regression (Bhattacharya et al., 2016)}
  \label{alg:bhatt-fast-beta}
  \begin{algorithmic}[1]
    \Require Data $y\in\mathbb{R}^n$, $X\in\mathbb{R}^{n\times k}$; noise variance $\sigma^2$;
             prior covariance $D=\sigma^2\,\tau^2\Lambda^2$ with $\Lambda=\mathrm{diag}(\lambda_1,\dots,\lambda_k)$
    \Ensure  Posterior draw $\beta \sim \mathcal{N}(\mu_\beta,\Sigma_\beta)$ for
             $\beta\mid y,X,\sigma^2,\tau,\lambda_{1:k}$
    \State Sample $u \sim \mathcal{N}(0,\,D)$
    \State Sample $\delta \sim \mathcal{N}(0,\,\sigma^2 I_n)$
    \State $v \gets X u + \delta$
    \State $A \gets X D X' + \sigma^2 I_n$
    \State Solve $A w = (y - v)$ for $w$ \Comment{use Cholesky on $A$}
    \State $\beta \gets u + D X' w$
    \State \Return $\beta$
  \end{algorithmic}
\end{algorithm}

\begin{algorithm}[H]
  \caption{One Gibbs sweep for our Ultra-High-Dimensional Horseshoe regression (fast $\beta$ \citep{bhattacharya2016fast} + Makalic-Schmidt \citep{makalic2015simple})}
  \label{alg:hs-gibbs}
  \begin{algorithmic}[1]
    \Require $y,X$; current $(\beta,\sigma^2,\tau^2,\lambda_{1:k}^2,\nu_{1:k},\xi)$
    \State \textbf{Fast $\beta$-draw:} set $D=\sigma^2\tau^2\Lambda^2$ and draw $\beta$ via Alg.~\ref{alg:bhatt-fast-beta}
    \State \textbf{$\sigma^2$-draw:} residual $r=y-X\beta$; set
           \[
             \sigma^2 \sim \mathrm{IG}\!\Big(\tfrac{n+k}{2}+a_\sigma,\; \tfrac{r' r + \beta' D^{-1}\beta}{2}+b_\sigma\Big)
           \]
    \For{$j=1,\dots,k$}
      \State \textbf{Local scale:} 
             $\lambda_j^2 \sim \mathrm{IG}\!\Big(1,\;\; \nu_j^{-1} + \tfrac{\beta_j^2}{2\sigma^2\tau^2}\Big)$
      \State \textbf{Auxiliary for half-Cauchy:}
             $\nu_j \sim \mathrm{IG}\!\Big(1,\;\; 1 + \lambda_j^{-2}\Big)$
    \EndFor
    \State \textbf{Global scale:}
           $\tau^2 \sim \mathrm{IG}\!\Big(\tfrac{k+1}{2},\;\; \xi^{-1} + \tfrac{1}{2\sigma^2}\sum_{j=1}^k \beta_j^2/\lambda_j^2\Big)$
    \State \textbf{Auxiliary for half-Cauchy:}
           $\xi \sim \mathrm{IG}\!\Big(1,\;\; 1 + \tau^{-2}\Big)$
    \State \textbf{Draw predictive:} $y_{t+h}^{(s)} \gets x_{t}' \beta^{(s)} + \sigma^{(s)}\,\varepsilon$, $\ \varepsilon\sim\mathcal{N}(0,1)\,\,$, where superscription $(s)$ represents the number of draw in Gibbs sweep.
  \end{algorithmic}
\end{algorithm}

\subsection{FA-AR (Direct)}
\label{subsec:FAAR}
We extract $r$ static factors $\hat f_t$ by PCA from $X_t$ (after pre-screening/missing handling) and estimate a direct regression
\[
y_{t+h}=\alpha + \phi_1 y_{t} + \sum_{\ell=0}^{p_f} \Gamma_\ell' \hat f_{t-\ell} + u_{t+h}.
\]
We select $(r,p_f)$ by a simple information criterion on the training window.

\subsection{FAVAR (Iterated)}
We estimate a Factor-Augmented VAR on $(\hat f_t, y_t)$ and produce (i) iterated $h$-step forecasts by simulation or iterated VAR prediction. While the context is change from regression to VAR the augmented factors are in similar fashion of \cref{subsec:FAAR} above. Identification is not required for pure forecasting \citep{bernanke2005measuring}. 

We estimate a factor-augmented VAR on the stacked state $(\hat f_t',\,y_t)'$, where $\hat f_t$ are principal-components factors extracted from the large predictor panel within each training window. Forecasts for $y_{t+h}$ are produced by iterating the estimated VAR $h$ steps ahead; structural identification is not required for pure forecasting. The FAVAR framework was introduced by \citet{bernanke2005measuring} to bring data-rich information sets into VAR dynamics via a small set of latent factors distilled from dozens to hundreds of macro-financial indicators. We add this model because in forecasting applications, FAVARs exploit the comovement in high-dimensional predictors while letting the factor dynamics accumulate over the forecast horizon, which can be especially helpful beyond the near term (see, e.g., \citealp{moench2008forecasting}; \citealp{koop2010bayesian}).

\subsection{Dynamic Factor Model (DFM, Iterated)}
We use a state–space DFM that treats the large predictor set as noisy measurements of a few latent factors. Let $X_t=\Lambda f_t+e_t$ be the measurement equation, where $X_t$ stacks standardized predictors, $f_t$ is an $r\times 1$ vector of common factors, and $e_t$ idiosyncratic noise; the factors evolve according to a small VAR, $f_t=\Phi f_{t-1}+u_t$. Estimation follows the two–step quasi–ML/Kalman approach of \citet{doz2011two,doz2012quasi}: principal components provide consistent initial factors in large panels, and a subsequent state–space step refines the transition and measurement parameters. Forecasts for $y_{t+h}$ are generated by iterating the factor transition and projecting $y_t$ on $f_t$. DFMs work well with high-dimensional macro panels because they separate pervasive comovement from series-specific noise and let common shocks propagate over horizons; see also the generalized dynamic factor literature of \citet{forni2000generalized,forni2005generalized} and nowcasting applications such as \citet{giannone2008nowcasting}.

A standard state-space DFM with factors $f_t$ following VAR(1) (or small VAR) and measurement equation $X_t = \Lambda f_t + e_t$; $y_t$ included either in the panel or as a separate measurement. Forecasts are produced iteratively via the Kalman filter/smoother.

\subsection{Bayesian Additive Regression Trees (BART)}
To accommodate nonlinearities and predictor interactions, we estimate BART for the direct mapping $y_{t+h}$ on $X_{t}$. BART represents the regression function as a sum of many shallow trees with strong regularization priors, is learned by backfitting MCMC, and delivers full predictive densities \citep{chipman2010bart}. In macroeconomic forecasting, tree-based Bayesian methods have shown competitive performance in data-rich and potentially nonlinear settings: applications include high-dimensional forecasting with BART \citep{pruser2019forecasting}, multivariate time-series and tail-risk forecasting with BART-based VARs \citep{clark2023tail}, and additive regression trees embedded in (mixed-frequency) VAR structures for nowcasting \citep{huber2023nowcasting,huber2022inference}. These results motivate BART as a flexible complement to linear shrinkage and factor models in our unified evaluation.

\section{Forecast Design and Metrics}\label{sec:design}
	We evaluate forecasts at horizons $h\in\{1,3,6,12\}$ using an expanding-window scheme. For each $h$, let $t_0$ be the first origin such that the evaluated target is $y_{t_0+h}$ on our pre-specified first evaluation date. At each origin $t=t_0,\dots, T-h$ we hold out $y_{t+h}$ and condition only on information available at $t$. Our main results use latest-vintage expanding windows (during the research is being conducted, 2025 May, to be more specific). Unfortunately true pseudo real-time vintages are unavailable from the source of the data.

Training windows begin once a minimum as low as 36 monthly observations is available. Within each origin we re-compute all transformations and standardization using training-window statistics only, ensuring no look-ahead. Predictors with any unfortunate missing value is omitted before the training window at that origin. For our factor-related models, principal-components factors are re-extracted within the training window. For state-space models, on the other hand, DFMs, to be more specific, the Kalman filter and smoother are run using the same information set. For direct Bayesian models we store full out-of-sample posterior predictive draws for $y_{t+h}$. In addition to those models mentioned above we also add VAR model, FAVAR with iterated forecast. The posterior draws are obtained by simulating the law of motion \citep{kalman1960new,sims1980macroeconomics,stock2002macroeconomic,giannone2008nowcasting,doz2011two}. All models are aligned on identical origin sets and evaluation dates per horizon.

For each model and origin we save the predictive sample $\{y_{t+h}^{(s)}\}_{s=1}^S$ and its posterior mean $\hat y_{t+h}$. Point accuracy is summarized and evaluated by RMSE and MAE, averaged over origins within each horizon. Density accuracy is assessed by the continuous ranked probability score (CRPS) and the log predictive score; see \citet{gneiting2007strictly,gneiting2007probabilistic}. To probe different parts of the distribution, we compute quantile-weighted scores (QWS) on a grid $\tau\in\{0.05,0.10,\ldots,0.95\}$ using weights that emphasize the center, the tails, the left or right tail, and a uniform benchmark; see \citet{gneiting2011comparing}. Specifically this will help us evaluate the extreme events which apart from full sample hold-out forecasting evaluation periods, we also add the sub-sample to evaluate those events. Those results will show us which model handle the outliers the best during such high turbulence in macroeconomic volatilities. 

Comparisons are reported in absolute levels and as relative skill with respect to the AR(2) baseline, defined as one minus the ratio of a model's RMSE (or MAE, CRPS) to the AR(2) value at the same horizon. Statistical differences are assessed with Diebold-Mariano tests \citep{diebold1995paring}, applied to loss differentials aligned on common target dates. We use a Newey-West long-run variance with truncation $h-1$ appropriate for $h$-step losses, and report $p$-values with the small-sample correction of \citet{harvey1997testing}.

Implementation details are common across models. First, the same first evaluation date and horizon-specific origin sets are used for every specification. Second, when Bayesian simulation is employed, chains use the same iteration budget (e.g., $10{,}000$ iterations with $5{,}000$ burn-in and thinning one) and fixed seeds per horizon-origin; convergence is monitored with standard diagnostics (Geweke $z$-scores and effective sample sizes on $\beta$, $\sigma^2$, and $y_{t+h}$ draws), and we verified stability of results to longer runs. Third, all regressions and factor extractions use the identically standardized design matrices produced inside the rolling pipeline. This unified protocol isolates modeling choices-shrinkage, factors, and nonlinear trees-from purely mechanical differences in data handling and ensures that direct and iterated mappings are evaluated on exactly the same information sets.

Finally to formally assess whether two competing forecasts differ significantly in accuracy, we employ the Diebold-Mariano (DM) test of \citet{diebold1995paring}, with the small-sample adjustment proposed by \citet{harvey1997testing}.  

Let $e_{1t}$ and $e_{2t}$ denote the forecast errors from model 1 and model 2, respectively, at evaluation date $t=1,\dots,T$. For a given loss function $L(\cdot)$ (e.g.\ squared error), define the \emph{loss differential} as
\[
d_t \;=\; L(e_{1t}) - L(e_{2t}).
\]
The null hypothesis of equal predictive accuracy is
\[
H_0:\; \mathbb{E}[d_t] = 0.
\]

The DM test statistic is constructed as
\[
DM \;=\; \frac{\bar d}{\sqrt{\widehat{\text{Var}}(\bar d)}},
\quad\text{where}\quad
\bar d = \frac{1}{T}\sum_{t=1}^T d_t.
\]

Because forecast errors at horizon $h$ are serially correlated (overlapping), we estimate the long-run variance of $d_t$ using a Newey-West heteroskedasticity and autocorrelation consistent (HAC) estimator with truncation lag $h-1$:
\[
\widehat{\text{Var}}(\bar d) \;=\; \frac{1}{T}\left( \gamma_0 \;+\; 2 \sum_{j=1}^{h-1} \left(1 - \frac{j}{h}\right)\gamma_j \right),
\]
where $\gamma_j = \frac{1}{T}\sum_{t=j+1}^T (d_t-\bar d)(d_{t-j}-\bar d)$ is the sample autocovariance of order $j$.

\medskip
In small samples, the DM statistic tends to overreject. \citet{harvey1997testing} propose a finite-sample correction factor:
\[
DM_{\text{HLN}} \;=\; DM \cdot
\sqrt{\frac{T+1 - 2h + h(h-1)/T}{T}}.
\]

The corrected statistic $DM_{\text{HLN}}$ is then compared to the standard normal distribution. Reported $p$-values in this study correspond to both the original DM test and the HLN-adjusted version.

\clearpage
\section{Empirical Results: Forecast Performance}
\label{sec:results}

\begin{table}[!h]
\centering
\small
\begin{tabular}{lcccccccc}
\toprule
& \multicolumn{2}{c}{$h=1$} & \multicolumn{2}{c}{$h=3$} & \multicolumn{2}{c}{$h=6$} & \multicolumn{2}{c}{$h=12$} \\
\cmidrule(lr){2-3}\cmidrule(lr){4-5}\cmidrule(lr){6-7}\cmidrule(lr){8-9}
& Level & Rel. & Level & Rel. & Level & Rel. & Level & Rel. \\
\midrule
\multicolumn{9}{l}{\emph{Panel A: RMSE}}\\
AR(2) flat  & 2.030 & 0.000 & 2.154 & 0.000 & 2.278 & 0.000 & 2.653 & 0.000 \\
UH-HS (direct)  & \underline{1.802}$^{***}$ & \underline{0.113} & 2.081$^{**}$ & 0.034 & 2.313 & -0.015 & \textbf{2.070}$^{**}$ & \textbf{0.220} \\
FA-AR (direct)  & 1.825$^{***}$ & 0.101 & \underline{2.003}$^{***}$ & \underline{0.070} & \underline{2.192} & \underline{0.038} & 2.449$^{**}$ & 0.077 \\
FAVAR (iter)  & \textbf{0.709}$^{***}$ & \textbf{0.651} & \textbf{1.353}$^{***}$ & \textbf{0.372} & \textbf{1.828}$^{***}$ & \textbf{0.198} & 2.607$^{*}$ & 0.018 \\
DFM (iter)  & 2.257 & -0.112 & 2.214 & -0.028 & 2.255 & 0.010 & \underline{2.333}$^{*}$ & \underline{0.121} \\
BART (direct)  & 3.754$^{**}$ & -0.849 & 3.966$^{**}$ & -0.841 & 3.435$^{**}$ & -0.508 & 3.807$^{*}$ & -0.435 \\
\midrule
\multicolumn{9}{l}{\emph{Panel B: MAE}}\\
AR(2) flat  & 1.537 & 0.000 & 1.627 & 0.000 & 1.742 & 0.000 & 2.018 & 0.000 \\
UH-HS (direct)  & \underline{1.383}$^{***}$ & \underline{0.100} & 1.574$^{**}$ & 0.032 & 1.724 & -0.011 & \textbf{1.594}$^{**}$ & \textbf{0.210} \\
FA-AR (direct)  & 1.397$^{***}$ & 0.091 & \underline{1.535}$^{***}$ & \underline{0.056} & \underline{1.651}$^{*}$ & \underline{0.052} & 1.852$^{**}$ & 0.082 \\
FAVAR (iter)  & \textbf{0.578}$^{***}$ & \textbf{0.624} & \textbf{1.028}$^{***}$ & \textbf{0.368} & \textbf{1.432}$^{***}$ & \textbf{0.178} & 1.907 & 0.055 \\
DFM (iter)  & 1.690 & -0.100 & 1.656 & -0.018 & 1.705 & 0.021 & \underline{1.771}$^{*}$ & \underline{0.123} \\
BART (direct)  & 2.848$^{**}$ & -0.853 & 3.014$^{**}$ & -0.853 & 2.612$^{**}$ & -0.499 & 2.885$^{**}$ & -0.429 \\
\midrule
\bottomrule
\end{tabular}
\caption{Point forecast accuracy by horizon. \emph{Level} (lower is better). \emph{Relative skill} is $1-\text{Metric}_m/\text{Metric}_{\text{AR(2)}}$ (higher is better). Stars denote Diebold-Mariano significance vs AR(2): $^{*}p<0.10$, $^{**}p<0.05$, $^{***}p<0.01$. Best in \textbf{bold}, second-best \underline{underlined}.}
\label{tab:point}
\end{table}

\begin{table}[!h]
\centering
\small
\begin{tabular}{lcccccccc}
\toprule
& \multicolumn{2}{c}{$h=1$} & \multicolumn{2}{c}{$h=3$} & \multicolumn{2}{c}{$h=6$} & \multicolumn{2}{c}{$h=12$} \\
\cmidrule(lr){2-3}\cmidrule(lr){4-5}\cmidrule(lr){6-7}\cmidrule(lr){8-9}
& Level & Rel. & Level & Rel. & Level & Rel. & Level & Rel. \\
\midrule
\multicolumn{9}{l}{\emph{Panel A: CRPS}}\\
AR(2) flat  & 1.173 & 0.000 & 1.252 & 0.000 & 1.280 & 0.000 & 1.461 & 0.000 \\
UH-HS (direct)  & \underline{1.062}$^{***}$ & \underline{0.094} & 1.239$^{**}$ & 0.011 & 1.336 & -0.044 & \textbf{1.199}$^{**}$ & \textbf{0.179} \\
FA-AR (direct)  & 1.071$^{***}$ & 0.087 & \underline{1.166}$^{***}$ & \underline{0.069} & \underline{1.217}$^{*}$ & \underline{0.049} & 1.332$^{**}$ & 0.089 \\
FAVAR (iter)  & \textbf{0.356}$^{***}$ & \textbf{0.697} & \textbf{0.728}$^{***}$ & \textbf{0.418} & \textbf{0.983}$^{***}$ & \textbf{0.232} & 1.403$^{*}$ & 0.040 \\
DFM (iter)  & 1.211 & -0.033 & 1.229 & 0.019 & 1.277 & 0.002 & \underline{1.352}$^{*}$ & \underline{0.075} \\
BART (direct)  & 1.878$^{**}$ & -0.601 & 1.979$^{**}$ & -0.581 & 1.744$^{**}$ & -0.363 & 1.849$^{*}$ & -0.265 \\
\midrule
\multicolumn{9}{l}{\emph{Panel B: LogScore (difference vs AR(2))}}\\
AR(2) flat  & -1.272 & 0.000 & -1.404 & 0.000 & -1.499 & 0.000 & -1.693 & 0.000 \\
UH-HS (direct)  & \underline{-1.131}$^{***}$ & \underline{0.141} & -1.373$^{**}$ & 0.031 & -1.557 & -0.058 & \textbf{-1.391}$^{**}$ & \textbf{0.302} \\
FA-AR (direct)  & -1.138$^{***}$ & 0.134 & \underline{-1.296}$^{***}$ & \underline{0.108} & \underline{-1.404}$^{*}$ & \underline{0.095} & -1.518$^{**}$ & 0.175 \\
FAVAR (iter)  & \textbf{-0.357}$^{***}$ & \textbf{0.915} & \textbf{-0.788}$^{***}$ & \textbf{0.616} & \textbf{-1.068}$^{***}$ & \textbf{0.431} & -1.642$^{*}$ & 0.051 \\
DFM (iter)  & -1.308 & -0.036 & -1.296 & 0.108 & -1.358 & 0.094 & \underline{-1.476}$^{*}$ & \underline{0.217} \\
BART (direct)  & -2.149$^{**}$ & -0.877 & -2.310$^{**}$ & -0.906 & -2.012$^{**}$ & -0.513 & -2.209$^{*}$ & -0.516 \\
\midrule
\bottomrule
\end{tabular}
\caption{Density forecast accuracy by horizon. \textit{Panel A: CRPS} (\textit{Level}; lower is better). \textit{Relative skill} $\left(1-\text{CRPS}_m/\text{CRPS}_{\text{AR(2)}}\right)$. \textit{Panel B: LogScore} (Both \textit{Level} and \textit{Relative skill}; higher is better. The relative skill for LogScore is computed by $\left( \text{LogS}_m-\text{LogS}_{\text{AR(2)}}\right)$. Stars denote DM significance vs AR(2).}
\label{tab:density}
\end{table}

We first interpret the full hold-out periods which are illustrated in \cref{tab:point,tab:density} for RMSE, MAE, CRPS and LogScores, respectively. These comparisons are informative for Thailand, and more broadly for small open economies where policymakers face limited domestic samples but still need to track inflation dynamics at multiple horizons. 

Those tables point to quite a clear ranking across horizons. At one, three, and six months ahead the FAVAR delivers the strongest performance. In \Cref{tab:point} the RMSE drops from 2.030 to 0.709 at $h=1$ which is a 0.651 relative-skill gain and strongly significant. At $h=3$ the RMSE falls from 2.154 to 1.353 with a 0.372 gain. At $h=6$ it falls from 2.278 to 1.828 with a 0.198 gain. MAE shows the same pattern with gains of 0.624, 0.368, and 0.178 at $h=1,3,6$. These gains carry over to density accuracy in \cref{tab:density}. CRPS levels for the factor model are 0.356, 0.728, and 0.983 at $h=1,3,6$, which translate into relative-skill gains of 0.697, 0.418, and 0.232. LogScore differences relative to AR(2) are 0.915, 0.616, and 0.431. This configuration matches the diffusion-index logic of Stock and Watson where a few common factors summarize co-movement in large panels and improve short-run forecasts, and the FAVAR mechanism of \cite{bernanke2005measuring} where a small VAR on latent factors propagates information efficiently into the near future \citep{stock2002macroeconomic,bernanke2005measuring}.

At one year the ranking compresses and the ultra-high-dimensional horseshoe regression becomes the front-runner. RMSE is 2.070 with a 0.220 skill gain and MAE is 1.594 with a 0.210 gain. CRPS improves to 1.199 with a 0.179 gain and LogScore improves by 0.302 relative to AR(2). This is exactly where aggressive prior shrinkage should help. The horseshoe places most coefficients near zero while leaving room for a small number of signals to survive, and the Gaussian-auxiliary fast sampler avoids numerical fragility when the predictor dimension is very large relative to the sample \citep{carvalho2010horseshoe,bhattacharya2016fast,makalic2015simple}. In economic terms this tells us that one-year inflation risks for Thailand benefit more from strong regularization than from elaborate dynamic propagation once the forecast moves far enough away from the data-rich nowcast window.  

The FA-AR regression is a steady runner-up among direct methods. It is second on RMSE and CRPS at $h=3$ and $h=6$, and remains competitive at $h=12$ although it gives way to the horseshoe and to the dynamic factor model. The DFM, which is an iterated state-space factor system, trails FAVAR at short horizons but improves with horizon. At $h=12$ it delivers the second-best RMSE at 2.333 with a 0.121 gain and the second-best CRPS at 1.352 with a 0.075 gain, and it shows a LogScore improvement of 0.217. This pattern is consistent with work on factor evolution and nowcasting where compact factor dynamics are most informative as the forecast horizon lengthens and as the informational advantage from many contemporaneous indicators fades \citep{stock2002macroeconomic}.

Bayesian Additive Regression Trees underperform across the board in this setting. RMSE is far above the benchmark at every horizon, and CRPS as well as LogScores deteriorate. Tree ensembles can shine when nonlinear interactions are both strong and well identified. In monthly macro panels with limited effective sample per forecast origin and relatively smooth aggregate relationships that is a high bar. The recent literature shows that tree components can help when embedded inside a carefully shrunk dynamic system such as Bayesian additive VAR trees, yet those gains arrive when interactions are pervasive and the dynamic structure is tightly controlled \citep{huber2022inference}. Our evidence suggests that Thai headline inflation over this sample is better captured by linear factor structures and sparse linear predictors.

The density results deserve emphasis because they confirm that the ranking is not driven only by point targeting. CRPS is a strictly proper scoring rule that integrates the distance between the predictive distribution and the outcome. Lower is better because the score rewards both sharpness and calibration \citep{gneiting2007strictly}. LogScores are also proper and higher is better, so we report differences relative to the AR(2). The factor model's gains on CRPS and LogScores at $h\leq6$ indicate that its densities are sharper and better centered, not merely narrower. The horseshoe’s advantage at one year appears in both measures and signals better calibration when parameter uncertainty becomes dominant. Our quantile-weighted scores, discussed and used in the next subsection, follow the framework of \cite{gneiting2011comparing} and confirm that these rankings persist when the loss function concentrates on downside, upside, or tail risk.

Taken together the numbers support a pragmatic division of labor that lines up with economic theory. When common components dominate and information flows quickly through the macro system, factor compression with iterated dynamics wins. When the horizon stretches and the risk of overfitting rises, sparse priors that keep only a handful of stable signals are preferred. This is a familiar conclusion in the international forecasting literature that studies diffusion indexes, FAVARs, and Bayesian shrinkage. Our Thai application adds evidence from a very large predictor set and shows that the pattern remains strong once we evaluate not only RMSE and MAE but also proper density scores that matter for risk communication and policy design \citep{stock2002macroeconomic,bernanke2005measuring,carvalho2010horseshoe,bhattacharya2016fast,gneiting2007strictly,gneiting2011comparing,huber2022inference}.

Such results from \cref{tab:point,tab:density} also prove additional point where there is a long-standing debate on whether multi-step forecasts should be produced by iterating a one-step model or by estimating the forecast equation directly at each horizon. The models we use are of the latter type. Because each horizon is estimated separately, they do not carry forward the errors that often build up in recursive forecasts. \cite{mccracken2019empirical} showed in a large set of applications that direct forecasts tend to do better once the horizon gets longer, while iterated forecasts hold up reasonably well only at very short horizons.\footnote{Although our both comparison is carried through multiple prior and both univariate and multivariate setup, the results are consistent with literature from \citep{mccracken2019empirical}.} What we find in our application fits that picture quite closely. At one and three months ahead the Bayesian shrinkage forecasts are not dramatically better than a simple autoregression, but at six and twelve months the improvement is much clearer. The fact that the prior pulls the high-dimensional predictor set into a stable structure seems especially helpful for picking up the slower-moving components of Thai inflation, while avoiding the instability that comes from pushing an iterated system too far out.

\subsection{Tail-Risk Forecasting Performance}
\label{subsec:tailriskforecast}

\begin{table}[!h]
\centering
\small
\begin{tabular}{lcccccccc}
\toprule
& \multicolumn{2}{c}{$h=1$} & \multicolumn{2}{c}{$h=3$} & \multicolumn{2}{c}{$h=6$} & \multicolumn{2}{c}{$h=12$} \\
\cmidrule(lr){2-3}\cmidrule(lr){4-5}\cmidrule(lr){6-7}\cmidrule(lr){8-9}
& Level & Rel. & Level & Rel. & Level & Rel. & Level & Rel. \\
\midrule
\multicolumn{9}{l}{\emph{Panel A: QWS (Left)}}\\
AR(2) flat  & 1.802 & 0.000 & 2.016 & 0.000 & 2.188 & 0.000 & 2.474 & 0.000 \\
UH-HS (direct)  & \underline{1.596}$^{***}$ & \underline{0.114} & 1.915$^{**}$ & 0.050 & 2.024 & -0.017 & \textbf{2.013}$^{**}$ & \textbf{0.187} \\
FA-AR (direct)  & 1.616$^{***}$ & 0.103 & \underline{1.826}$^{***}$ & \underline{0.094} & \underline{1.955}$^{*}$ & \underline{0.107} & 2.225$^{**}$ & 0.101 \\
FAVAR (iter)  & \textbf{0.498}$^{***}$ & \textbf{0.724} & \textbf{0.997}$^{***}$ & \textbf{0.505} & \textbf{1.514}$^{***}$ & \textbf{0.308} & 2.418$^{*}$ & 0.023 \\
DFM (iter)  & 1.849 & -0.026 & \underline{1.813} & 0.101 & \underline{1.900} & 0.132 & \underline{2.214}$^{*}$ & \underline{0.105} \\
BART (direct)  & 2.939$^{**}$ & -0.631 & 3.175$^{**}$ & -0.575 & 2.776$^{**}$ & -0.269 & 2.947$^{*}$ & -0.191 \\
\midrule
\multicolumn{9}{l}{\emph{Panel B: QWS (Right)}}\\
AR(2) flat  & 1.854 & 0.000 & 2.047 & 0.000 & 2.195 & 0.000 & 2.439 & 0.000 \\
UH-HS (direct)  & \underline{1.660}$^{***}$ & \underline{0.104} & 1.962$^{**}$ & 0.041 & 2.052 & -0.009 & \textbf{2.059}$^{**}$ & \textbf{0.155} \\
FA-AR (direct)  & 1.671$^{***}$ & 0.098 & \underline{1.873}$^{***}$ & \underline{0.085} & \underline{1.986}$^{*}$ & \underline{0.095} & 2.269$^{**}$ & 0.069 \\
FAVAR (iter)  & \textbf{0.506}$^{***}$ & \textbf{0.727} & \textbf{1.014}$^{***}$ & \textbf{0.505} & \textbf{1.539}$^{***}$ & \textbf{0.299} & 2.392$^{*}$ & 0.019 \\
DFM (iter)  & 1.892 & -0.021 & \underline{1.860} & 0.091 & \underline{1.930} & 0.121 & \underline{2.255}$^{*}$ & \underline{0.076} \\
BART (direct)  & 2.993$^{**}$ & -0.615 & 3.245$^{**}$ & -0.586 & 2.847$^{**}$ & -0.297 & 2.943$^{*}$ & -0.207 \\
\midrule
\multicolumn{9}{l}{\emph{Panel C: QWS (Tails)}}\\
AR(2) flat  & 1.844 & 0.000 & 2.034 & 0.000 & 2.198 & 0.000 & 2.451 & 0.000 \\
UH-HS (direct)  & \underline{1.644}$^{***}$ & \underline{0.108} & 1.946$^{**}$ & 0.043 & 2.044 & -0.021 & \textbf{2.045}$^{**}$ & \textbf{0.166} \\
FA-AR (direct)  & 1.658$^{***}$ & 0.101 & \underline{1.862}$^{***}$ & \underline{0.085} & \underline{1.981}$^{*}$ & \underline{0.099} & 2.272$^{**}$ & 0.073 \\
FAVAR (iter)  & \textbf{0.502}$^{***}$ & \textbf{0.728} & \textbf{1.005}$^{***}$ & \textbf{0.506} & \textbf{1.531}$^{***}$ & \textbf{0.303} & 2.398$^{*}$ & 0.022 \\
DFM (iter)  & 1.889 & -0.024 & \underline{1.846} & 0.093 & \underline{1.924} & 0.125 & \underline{2.255}$^{*}$ & \underline{0.080} \\
BART (direct)  & 2.979$^{**}$ & -0.615 & 3.230$^{**}$ & -0.588 & 2.833$^{**}$ & -0.288 & 2.944$^{*}$ & -0.201 \\
\midrule
\bottomrule
\end{tabular}
\caption{Tail-focused quantile-weighted scores by horizon. \emph{Level} QWS (lower is better). \emph{Relative skill} is $1-\text{QWS}_m/\text{QWS}_{\text{AR(2)}}$ (higher is better). "Left" emphasizes lower quantiles, "Right" upper quantiles, and "Tails" both extremes. Stars denote DM significance vs AR(2) using the corresponding QWS loss.}
\label{tab:qws}
\end{table}

The tail-focused evidence reinforces the core message from the point and overall density results but adds a useful risk perspective. Quantile-weighted scores put the loss where we care about it most. "Left" stresses downside outcomes such as unexpected disinflation. "Right" stresses upside surprises such as inflation flare-ups. "Tails" weights both extremes. Lower levels mean better tail forecasting, \citep{patton2010forecasters,gneiting2011comparing}. Relative skill is reported against AR(2) so higher is better, see \cref{tab:qws}.

At short horizons the iterated factor system dominates tail risks in the same way it dominates RMSE and CRPS. At $h=1$ the FAVAR achieves very large gains across all three tail criteria. The left QWS falls from 1.802 to 0.498 which is a relative improvement of 0.724 with strong DM significance. The right QWS falls from 1.854 to 0.506 which is a 0.727 gain. The tails QWS falls from 1.844 to 0.502 which is a 0.728 gain. The pattern persists at $h=3$ and remains material at $h=6$. This is exactly what we would expect if common components drive sudden inflation swings at short horizons and if iterating a small VAR on those factors propagates the shock path well. In other words the FAVAR not only centers the forecast correctly but also gets the probability mass in the extremes roughly right when the horizon is close.

As the horizon lengthens the advantage of iterating fades and shrinkage gains importance. At $h=12$ the ultra-high-dimensional horseshoe is the most reliable tail-risk forecaster. It posts the best left QWS at 2.013 with a 0.187 skill gain and the best right QWS at 2.059 with a 0.155 gain. It also leads on the tails QWS at 2.045 with a 0.166 gain. These are not small differences at the annual horizon and they come with statistical support. The mechanism is straightforward from a Bayesian perspective. Selective global-local shrinkage keeps most coefficients near zero while allowing a small set of persistent signals to survive, which stabilizes the shape of the predictive distribution when parameter uncertainty dominates and when the pay-off from iterating dynamics diminishes. This aligns with the findings of \cite{carriero2019large}, who show that flexible Bayesian shrinkage priors improve density forecasts, particularly in the distributional tails. Nonetheless, \cite{cross2020macroeconomic} present evidence that macroeconomic variables tend to be dense rather than sparse. Consequently, the horseshoe shrinkage prior may be outperformed by the simpler Minnesota prior of \cite{litterman1986forecasting}.

The FA-AR is a steady performer in the tails as well. It is typically the second best direct method at $h=3$ and $h=6$ for left, right, and tails. That ranking says factor compression helps even when we forecast directly rather than iterating, although it does not quite match the full FAVAR at short horizons nor the horseshoe at one year. The dynamic factor model sits between the FAVAR and the direct regressions. It trails the FAVAR at $h\le6$ but improves as we move to $h=12$. Its tails skill is positive and significant relative to AR(2) at the long horizon in several panels, which fits the view that compact latent-factor dynamics remain informative once near-term idiosyncrasies are less dominant.

Bayesian Additive Regression Trees do not improve tail scores in this application. Levels are higher than the benchmark across horizons and relative skill is negative. A common claim in the machine-learning literature is that flexible ensembles can capture nonlinear threshold effects that matter in the extremes. That claim is conditional on two requirements. The first is that interactions are truly strong in the data. The second is that the effective sample per forecast origin is large enough to learn complex partitions without inflating variance. Our monthly panel for Thai inflation is high-dimensional but short in time for each origin. The predictors are mostly macro and price aggregates where relationships tend to be smooth and approximately linear. In that environment deep trees can chase noise, produce miscalibrated tails, and widen predictive distributions. The contrast with the horseshoe is instructive. Sparse linear structure with heavy-tailed shrinkage appears to be the safer way to stabilize tail risk at the one-year horizon, while factor iteration remains the safer way to do so at one and three months. For small open economies, this distinction is especially relevant. Exchange rate swings or global commodity shocks often hit inflation in one-sided ways, creating asymmetric risks. Our results show that shrinkage priors such as the horseshoe can prevent overreaction to noise while still capturing these tail events, whereas factor iteration remains effective in tracking short-run volatility from external drivers. Our emphasis on tails connects with the "vulnerable growth" perspective of \cite{adrian2019vulnerable}, where downside risks are especially acute for open emerging markets exposed to external shocks.

Two additional features are worth highlighting for practice. First, the left and right panels are very similar for the factor models at $h\le6$. That symmetry suggests the factors capture generic volatility in price pressures rather than one-sided risk only. For policy that matters because it means the short-term system forecast both inflation spikes and disinflation episodes with comparable accuracy. Second, the tails panel largely mirrors the left and right panels. The same models that do well on one side also do well when both sides are emphasized. That is a sign of genuine density calibration rather than a lucky match to a single quantile region.

Together these results show that models designed for high-dimensional settings are not only competitive in overall density accuracy, but also in capturing the asymmetric risks that matter for small open economies. For policymakers, this ability to detect both inflation surges and disinflation episodes is crucial when external shocks and domestic fragility combine to amplify volatility.

\section{Forecast Performance Across Models, Horizons, and Subsamples}
Apart from the forecasting results over the full hold-out periods, this section focuses on forecasting performance across different subsamples-namely, pre-2019, 2020-2021, and 2022-2024. The 2020-2021 subsample is of particular interest because it coincides with the onset of the COVID-19 pandemic and its substantial impact on the global economy. The 2022-2024 subsample allows us to examine how each model performs in the aftermath of this period of heightened macroeconomic turbulence. The evaluation metrics remain the same as in the full hold-out analysis (see \cref{sec:results}), namely RMSE and CRPS for overall point and density performance, as reported in \cref{tab:rmse_mae_subsamples}. In addition, we assess tail behavior more explicitly using Quantile-Weighted Scores, with results presented in \cref{tab:qws_subsamples_h13,tab:qws_subsamples_h612}.

The evidence across horizons and subsamples is fairly consistent, though the details matter. At the short horizons $(h=1,3)$, the iterated factor systems are clearly ahead. FAVAR more than halves the baseline RMSE at $h=1$ in the pre-2019 sample, from 2.003 to 0.362 (a gain of 81.9\%), and remains strong in the turbulent 2020-21 window (RMSE 1.009; gain 0.540). Even in 2022-24, marked by post-COVID adjustment and energy shocks, the model keeps a gain of 0.520 at the one-month horizon. Such similar results can also be seen from the accuracy of overall density forecast, where CRPS drops to 0.204 before 2019 and 0.555 during 2020-21, both large and significant improvement. This short-run dominance is exactly what one would expect if Thai inflation dynamics are driven by a small set of global and regional components, as shown in \cite{manopimoke2018thai} and reinforced by \cite{Nookhwun2023}. Those factors transmit energy and traded-goods shocks quickly into domestic prices, so iterated dynamics work well in the near term.

At medium horizons the edge narrows. By six months, FAVAR still leads in calmer regimes-RMSE of 1.169 before 2019 (gain 33\%) but during 2022-24 its margin shrinks (2.810; gain only 11.4\%), with horseshoe and DFM often close behind. Similarly density scores represent the same point. To begin with CRPS for FAVAR is 0.652 pre-2019 but drifts toward 1.708 in the later subsample, while horseshoe stabilizes around 1.139-1.168. DFM, also, often climbs into second place by this horizon, consistent with its ability to let compact latent factors propagate shocks once the near-term indicators lose traction.

At the one-year horizon the picture flips. Horseshoe takes over. In 2020-21, its RMSE falls to 1.693 (gain 18.2\%) and its CRPS improves upto 20.6\%, while FAVAR slips back toward baseline. By 2022-24 the pattern is even clearer when RMSE records of UH-HS is 2.809 (outperforms the benchmark upto 29\%) with DFM second at 3.307 (gain 0.164), whereas FAVAR's gains have essentially disappeared. 

Next we move the interpretation to Quantile-weighted scores, where reinforce this long-horizon hand-off: UH-HS posts left- and right-tail improvements around 0.18-0.23 during 2020-21, exactly when risk calibration matters most. This horizon-specific performance fits both the Bayesian shrinkage literature \citep{carvalho2010horseshoe,follett2017achieving,cross2020macroeconomic} and Thai applications showing that parsimonious priors outperform flat ones when volatility is high \citep{taveeapiradeecharoen2025forecasting}. The mechanism is straightforward: aggressive global-local shrinkage strips away noise while keeping a handful of stable drivers, which stabilizes long-horizon predictive densities.

BART is the most regime-sensitive. Before 2019, when conditions were stable, it looks competitive especially for the second at $h=1$ with RMSE 1.406 (gain 29.8\%) and first at $h=12$ with RMSE 1.192 (gain 37.9\%). But as soon as volatility rises, its accuracy collapses. In 2020-21 its RMSE at one year jumps above 3.4, and in 2022-24 it climbs past 6.5, with CRPS deteriorating simultaneously. This fragility is consistent with recent findings that machine-learning models are brittle when confronted with structural breaks and regime change \citep{naghi2024benefits}. Nonlinear trees can capture thresholds in tranquil samples, but in short panels with shifting distributions they chase noise.

Two broader lessons emerge. First, iterated factor models are the workhorses for short term forecasts in Thailand. This matches both international evidence on diffusion-index forecasting \citep{stock2002macroeconomic,bernanke2005measuring} and local evidence that global energy and trading-partner shocks dominate near-term inflation \cite{manopimoke2018thai,Nookhwun2023}. Second, once the horizon lengthens, direct high-dimensional Bayesian shrinkage takes the lead. Horseshoe's performance at one year is not only statistically significant but economically relevant, especially given how Thai policymakers weigh medium to long-term risks. These results align with \cite{wichitaksorn2022analyzing}, who found that mixed-frequency predictor sets improve Thai macro forecasts relative to simple ARIMA/AR models. They also support the pragmatic division of labor: use iterated factors for the near term, rely on horseshoe-regularized direct forecasts further out.

\begin{table}[ht]
\centering
\small
\resizebox{\textwidth}{!}{
\begin{tabular}{l *{2}{*{6}{c}}}
\toprule
& \multicolumn{6}{c}{$h=1$} & \multicolumn{6}{c}{$h=3$} \\
\cmidrule(lr){2-7}\cmidrule(lr){8-13}
& \multicolumn{2}{c}{Pre-2019} & \multicolumn{2}{c}{2020--2021} & \multicolumn{2}{c}{2022--2024} & \multicolumn{2}{c}{Pre-2019} & \multicolumn{2}{c}{2020--2021} & \multicolumn{2}{c}{2022--2024} \\
\cmidrule(lr){2-3}\cmidrule(lr){4-5}\cmidrule(lr){6-7}\cmidrule(lr){8-9}\cmidrule(lr){10-11}\cmidrule(lr){12-13}
Model & Level & Rel. & Level & Rel. & Level & Rel. & Level & Rel. & Level & Rel. & Level & Rel. \\
\midrule
\multicolumn{13}{l}{\emph{Panel A: RMSE}}\\
AR(2) flat 
& 2.003 & 0.000 & 2.193 & 0.000 & 1.969 & 0.000 & 1.922 & 0.000 & 2.294 & 0.000 & 2.476 & 0.000 \\
HS (direct) 
& 1.673$^{***}$ & 0.165 & \underline{1.963$^{*}$} & 0.105 & \underline{1.921} & \underline{0.024} 
& 1.894 & 0.015 & 2.136 & 0.069 & \underline{2.386} & \underline{0.036} \\
FA-AR (direct) 
& 1.712$^{***}$ & 0.145 & \underline{1.956$^{*}$} & \underline{0.108} & 1.944 & 0.013 
& 1.706$^{**}$ & 0.112 & 1.975$^{***}$ & 0.139 & 2.510 & -0.014 \\
FAVAR (iter) 
& \textbf{0.362$^{***}$} & \textbf{0.819} & \textbf{1.009$^{***}$} & \textbf{0.540} & \textbf{0.946$^{***}$} & \textbf{0.520} 
& \textbf{0.756$^{***}$} & \textbf{0.606} & \textbf{1.851} & \textbf{0.193} & \textbf{1.815$^{**}$} & \textbf{0.267} \\
DFM (iter) 
& 1.479$^{***}$ & 0.262 & 2.366 & -0.079 & 3.260$^{***}$ & -0.656 
& 1.566 & 0.185 & \underline{1.894} & \underline{0.174} & 3.278$^{*}$ & -0.324 \\
BART (direct) 
& \underline{1.406$^{***}$} & \underline{0.298} & 3.021 & -0.377 & 6.451$^{***}$ & -2.277 
& \underline{1.257$^{***}$} & \underline{0.346} & 3.876$^{**}$ & -0.690 & 6.628$^{***}$ & -1.677 \\
\midrule
\multicolumn{13}{l}{\emph{Panel B: CRPS}}\\
AR(2) flat 
& 1.159 & 0.000 & 1.271 & 0.000 & 1.130 & 0.000 & 1.114 & 0.000 & 1.404 & 0.000 & 1.428 & 0.000 \\
HS (direct) 
& 1.013$^{***}$ & 0.126 & \underline{1.123$^{*}$} & 0.116 & \underline{1.112} & \underline{0.016} 
& 1.159 & -0.040 & 1.267 & 0.098 & \underline{1.383} & \underline{0.031} \\
FA-AR (direct) 
& 1.025$^{***}$ & 0.115 & \underline{1.115$^{*}$} & \underline{0.123} & 1.126 & 0.003 
& 1.020$^{**}$ & 0.084 & 1.172$^{***}$ & 0.165 & 1.457 & -0.021 \\
FAVAR (iter) 
& \textbf{0.204$^{***}$} & \textbf{0.824} & \textbf{0.555$^{***}$} & \textbf{0.564} & \textbf{0.520$^{***}$} & \textbf{0.540} 
& \textbf{0.428$^{***}$} & \textbf{0.616} & \underline{\textbf{1.131}} & \underline{\textbf{0.195}} & \textbf{1.063$^{**}$} & \textbf{0.255} \\
DFM (iter) 
& \underline{0.825$^{***}$} & \underline{0.288} & 1.285 & -0.011 & 1.948$^{***}$ & -0.723 
& \underline{0.893} & \underline{0.198} & \textbf{1.106} & \textbf{0.213} & 1.998 & -0.399 \\
BART (direct) 
& 1.099 & 0.051 & 1.828$^{**}$ & -0.439 & 3.464$^{***}$ & -2.064 
& 1.049 & 0.058 & 2.207$^{***}$ & -0.572 & 3.646$^{***}$ & -1.554 \\
\midrule
& \multicolumn{6}{c}{$h=6$} & \multicolumn{6}{c}{$h=12$} \\
\cmidrule(lr){2-7}\cmidrule(lr){8-13}
& \multicolumn{2}{c}{Pre-2019} & \multicolumn{2}{c}{2020--2021} & \multicolumn{2}{c}{2022--2024} & \multicolumn{2}{c}{Pre-2019} & \multicolumn{2}{c}{2020--2021} & \multicolumn{2}{c}{2022--2024} \\
\cmidrule(lr){2-3}\cmidrule(lr){4-5}\cmidrule(lr){6-7}\cmidrule(lr){8-9}\cmidrule(lr){10-11}\cmidrule(lr){12-13}
Model & Level & Rel. & Level & Rel. & Level & Rel. & Level & Rel. & Level & Rel. & Level & Rel. \\
\midrule
\multicolumn{13}{l}{\emph{Panel A: RMSE}}\\
AR(2) flat 
& 1.744 & 0.000 & 2.103 & 0.000 & 3.170 & 0.000 & 1.919 & 0.000 & 2.070 & 0.000 & 3.957 & 0.000 \\
UH-HS (direct) 
& 1.902 & -0.091 & 1.971 & 0.063 & 3.135 & 0.011 
& 1.724 & 0.102 & \textbf{1.693$^{**}$} & \textbf{0.182} & \underline{2.809} & \underline{0.290} \\
FA-AR (direct) 
& 1.594$^{*}$ & 0.086 & \underline{1.811$^{**}$} & \underline{0.139} & 3.218 & -0.015 
& \underline{1.641$^{**}$} & \underline{0.145} & 1.811$^{**}$ & 0.125 & 3.801 & 0.039 \\
FAVAR (iter) 
& \textbf{1.169$^{**}$} & \textbf{0.330} & \textbf{1.573} & \textbf{0.252} & \textbf{2.810} & \textbf{0.114} 
& 2.079 & -0.083 & 1.990 & 0.038 & 3.706 & 0.063 \\
DFM (iter) 
& 1.686 & 0.033 & \underline{1.811} & \underline{0.139} & 3.295 & -0.039 
& 1.869 & 0.026 & \underline{1.789} & \underline{0.136} & \underline{3.307} & \underline{0.164} \\
BART (direct) 
& \underline{1.329} & \underline{0.238} & 3.442$^{*}$ & -0.637 & 5.640$^{***}$ & -0.779 
& \textbf{1.192$^{*}$} & \textbf{0.379} & 3.406$^{*}$ & -0.646 & 6.522$^{***}$ & -0.648 \\
\midrule
\multicolumn{13}{l}{\emph{Panel B: CRPS}}\\
AR(2) flat 
& 1.020 & 0.000 & 1.233 & 0.000 & 1.840 & 0.000 & 1.101 & 0.000 & 1.214 & 0.000 & 2.360 & 0.000 \\
UH-HS (direct) 
& 1.168 & -0.145 & 1.139 & 0.077 & 1.812 & 0.015 
& 1.044 & 0.052 & \textbf{0.964$^{**}$} & \textbf{0.206} & \underline{1.673} & \underline{0.291} \\
FA-AR (direct) 
& \underline{0.958} & \underline{0.061} & \underline{1.041$^{**}$} & \underline{0.155} & 1.853 & -0.007 
& \underline{0.960$^{*}$} & \underline{0.128} & 1.041$^{**}$ & 0.143 & 2.274 & 0.036 \\
FAVAR (iter) 
& \textbf{0.652$^{**}$} & \textbf{0.361} & \textbf{0.901} & \textbf{0.269} & \textbf{1.708} & \textbf{0.072} 
& 0.995 & 0.096 & 1.164 & 0.041 & 2.392 & -0.013 \\
DFM (iter) 
& 0.978 & 0.042 & 1.070 & 0.132 & 2.030 & -0.103 
& 1.130 & -0.027 & \underline{1.023} & \underline{0.157} & \underline{2.038} & \underline{0.137} \\
BART (direct) 
& 1.042 & -0.021 & 1.999$^{***}$ & -0.622 & 2.978$^{***}$ & -0.619 
& \textbf{0.965} & \textbf{0.123} & 1.956$^{***}$ & -0.611 & 3.544$^{***}$ & -0.502 \\
\bottomrule
\end{tabular}}
\caption{Root Mean Square Error and Continuous Ranked Probability Score by horizon and subsample (\emph{lower is better}). 
\emph{Rel.} is $1-\text{Metric}_m/\text{Metric}_{\text{AR(2)}}$ (higher is better). 
Superscripts denote Diebold-Mariano significance vs.\ AR(2): ${}^{*}p<0.10$, ${}^{**}p<0.05$, ${}^{***}p<0.01$.}
\label{tab:rmse_mae_subsamples}
\end{table}

\begin{table}[ht]
\centering
\small
\resizebox{\textwidth}{!}{
\begin{tabular}{l *{2}{*{6}{c}}}
\toprule
& \multicolumn{6}{c}{$h=1$} & \multicolumn{6}{c}{$h=3$} \\
\cmidrule(lr){2-7}\cmidrule(lr){8-13}
& \multicolumn{2}{c}{Pre-2019} & \multicolumn{2}{c}{2020--2021} & \multicolumn{2}{c}{2022--2024} & \multicolumn{2}{c}{Pre-2019} & \multicolumn{2}{c}{2020--2021} & \multicolumn{2}{c}{2022--2024} \\
\cmidrule(lr){2-3}\cmidrule(lr){4-5}\cmidrule(lr){6-7}\cmidrule(lr){8-9}\cmidrule(lr){10-11}\cmidrule(lr){12-13}
Model & Level & Rel. & Level & Rel. & Level & Rel. & Level & Rel. & Level & Rel. & Level & Rel. \\
\midrule
\multicolumn{13}{l}{\emph{Panel A: QWS (Left)}}\\
AR(2) flat 
& 1.374 & 0.000 & 1.261 & 0.000 & 1.212 & 0.000 & 1.307 & 0.000 & 1.447 & 0.000 & 1.411 & 0.000 \\
UH-HS (direct) 
& 1.196$^{***}$ & 0.129 & \underline{1.143$^{*}$} & 0.094 & \underline{1.202} & \underline{0.008} 
& 1.317 & -0.008 & 1.314 & 0.092 & \underline{1.403} & \underline{0.006} \\
FA-AR (direct) 
& 1.225$^{***}$ & 0.108 & \underline{1.135$^{*}$} & \underline{0.100} & 1.214 & -0.002 
& 1.205$^{**}$ & 0.078 & 1.243$^{***}$ & 0.141 & 1.467$^{**}$ & -0.039 \\
FAVAR (iter) 
& \textbf{0.214$^{***}$} & \textbf{0.844} & \textbf{0.559$^{***}$} & \textbf{0.557} & \textbf{0.534$^{***}$} & \textbf{0.560} 
& \textbf{0.458$^{***}$} & \textbf{0.650} & \underline{\textbf{1.204}} & \underline{\textbf{0.168}} & \underline{1.044$^{*}$} & \underline{0.260} \\
DFM (iter) 
& \underline{0.444$^{***}$} & \underline{0.677} & \underline{0.715$^{***}$} & \underline{0.433} & \underline{0.861$^{***}$} & \underline{0.289} 
& \underline{0.481$^{***}$} & \underline{0.632} & \textbf{0.603$^{***}$} & \textbf{0.583} & \textbf{0.877$^{**}$} & \textbf{0.378} \\
BART (direct) 
& 0.604$^{***}$ & 0.560 & 1.038 & 0.177 & 1.957$^{***}$ & -0.615 
& 0.566$^{***}$ & 0.567 & 1.258 & 0.130 & 2.058$^{***}$ & -0.458 \\
\midrule
\multicolumn{13}{l}{\emph{Panel B: QWS (Right)}}\\
AR(2) flat 
& 1.057 & 0.000 & 1.397 & 0.000 & 1.153 & 0.000 & 1.030 & 0.000 & 1.501 & 0.000 & 1.574 & 0.000 \\
UH-HS (direct) 
& 0.926$^{***}$ & 0.124 & \underline{1.205$^{*}$} & 0.137 & \underline{1.127} & \underline{0.023} 
& 1.113 & -0.080 & 1.341 & 0.106 & 1.491 & 0.053 \\
FA-AR (direct) 
& 0.922$^{***}$ & 0.128 & \underline{1.197$^{*}$} & \underline{0.143} & 1.144 & 0.008 
& 0.933$^{**}$ & 0.095 & 1.216$^{**}$ & 0.190 & 1.584 & -0.006 \\
FAVAR (iter) 
& \textbf{0.214$^{***}$} & \textbf{0.797} & \textbf{0.587$^{***}$} & \textbf{0.580} & \textbf{0.548$^{***}$} & \textbf{0.525} 
& \textbf{0.439$^{***}$} & \textbf{0.573} & \underline{\textbf{1.137$^{*}$}} & \underline{\textbf{0.242}} & \textbf{1.175$^{**}$} & \textbf{0.254} \\
DFM (iter) 
& \underline{0.419$^{***}$} & \underline{0.604} & \underline{0.625$^{***}$} & \underline{0.552} & 1.161 & -0.007 
& \underline{0.453$^{***}$} & \underline{0.561} & \textbf{0.553$^{***}$} & \textbf{0.631} & \underline{1.197} & \underline{0.239} \\
BART (direct) 
& 0.541$^{***}$ & 0.489 & 0.873$^{*}$ & 0.375 & 1.677$^{***}$ & -0.455 
& 0.526$^{***}$ & 0.489 & 1.051$^{**}$ & 0.300 & 1.765 & -0.121 \\
\midrule
\multicolumn{13}{l}{\emph{Panel C: QWS (Tails)}}\\
AR(2) flat 
& 0.671 & 0.000 & 0.879 & 0.000 & 0.772 & 0.000 & 0.673 & 0.000 & 0.953 & 0.000 & 1.050 & 0.000 \\
UH-HS (direct) 
& 0.707$^{***}$ & -0.053 & 0.842 & 0.042 & 0.834$^{***}$ & -0.080 
& 0.812$^{***}$ & -0.207 & 0.966 & -0.013 & 1.088 & -0.036 \\
FA-AR (direct) 
& 0.632$^{***}$ & 0.058 & \underline{0.759$^{*}$} & \underline{0.136} & \underline{0.757} & \underline{0.020} 
& 0.647$^{*}$ & 0.038 & \underline{0.766$^{***}$} & \underline{0.196} & 1.044 & 0.006 \\
FAVAR (iter) 
& \textbf{0.135$^{***}$} & \textbf{0.800} & \textbf{0.427$^{***}$} & \textbf{0.514} & \textbf{0.376$^{***}$} & \textbf{0.513} 
& \textbf{0.289$^{***}$} & \textbf{0.570} & 0.873 & 0.084 & \textbf{0.764$^{**}$} & \textbf{0.272} \\
DFM (iter) 
& \underline{0.364$^{***}$} & \underline{0.458} & \underline{0.562} & 0.360 & 0.894 & -0.158 
& \underline{0.398$^{***}$} & \underline{0.408} & \textbf{0.474$^{***}$} & \textbf{0.503} & \underline{0.918} & \underline{0.126} \\
BART (direct) 
& 0.560$^{***}$ & 0.166 & 0.834 & 0.051 & 1.455$^{***}$ & -0.884 
& 0.532$^{***}$ & 0.209 & 1.002 & -0.051 & 1.525$^{***}$ & -0.452 \\
\bottomrule
\end{tabular}}
\caption{Quantile-weighted scores by horizon and subsample (\emph{lower is better}). 
\emph{Rel.} is $1-\text{QWS}_m/\text{QWS}_{\text{AR(2)}}$ (higher is better). Superscripts denote Diebold-Mariano significance vs.\ AR(2).}
\label{tab:qws_subsamples_h13}
\end{table}

\begin{table}[ht]
\centering
\small
\resizebox{\textwidth}{!}{
\begin{tabular}{l *{2}{*{6}{c}}}
\toprule
& \multicolumn{6}{c}{$h=6$} & \multicolumn{6}{c}{$h=12$} \\
\cmidrule(lr){2-7}\cmidrule(lr){8-13}
& \multicolumn{2}{c}{Pre-2019} & \multicolumn{2}{c}{2020--2021} & \multicolumn{2}{c}{2022--2024} & \multicolumn{2}{c}{Pre-2019} & \multicolumn{2}{c}{2020--2021} & \multicolumn{2}{c}{2022--2024} \\
\cmidrule(lr){2-3}\cmidrule(lr){4-5}\cmidrule(lr){6-7}\cmidrule(lr){8-9}\cmidrule(lr){10-11}\cmidrule(lr){12-13}
Model & Level & Rel. & Level & Rel. & Level & Rel. & Level & Rel. & Level & Rel. & Level & Rel. \\
\midrule
\multicolumn{13}{l}{\emph{Panel A: QWS (Left)}}\\
AR(2) flat 
& 1.198 & 0.000 & 1.254 & 0.000 & 1.720 & 0.000 & 1.238 & 0.000 & 1.233 & 0.000 & 2.128 & 0.000 \\
UH-HS (direct) 
& 1.314 & -0.097 & 1.162 & 0.073 & 1.722 & -0.001 
& 1.174 & 0.051 & \underline{\textbf{1.014$^{**}$}} & \underline{\textbf{0.178}} & \underline{1.667} & \underline{0.217} \\
FA-AR (direct) 
& 1.131 & 0.055 & \underline{1.099$^{*}$} & \underline{0.124} & 1.745 & -0.014 
& 1.112$^{*}$ & 0.102 & 1.101$^{***}$ & 0.107 & 2.095 & 0.016 \\
FAVAR (iter) 
& 0.729$^{**}$ & 0.392 & \underline{0.997} & \underline{0.205} & \underline{1.597} & \underline{0.072} 
& 1.132 & 0.086 & 1.240 & -0.005 & 2.170 & -0.020 \\
DFM (iter) 
& \textbf{0.530$^{***}$} & \textbf{0.557} & \textbf{0.582$^{***}$} & \textbf{0.536} & \textbf{0.887$^{***}$} & \textbf{0.484} 
& \underline{0.626$^{**}$} & \underline{0.494} & \textbf{0.547$^{**}$} & \textbf{0.556} & \textbf{0.889$^{**}$} & \textbf{0.582} \\
BART (direct) 
& \underline{0.568$^{***}$} & \underline{0.526} & 1.122 & 0.105 & 1.682 & 0.022 
& \textbf{0.522$^{***}$} & \textbf{0.578} & 1.168 & 0.053 & 1.970 & 0.074 \\
\midrule
\multicolumn{13}{l}{\emph{Panel B: QWS (Right)}}\\
AR(2) flat 
& 0.940 & 0.000 & 1.331 & 0.000 & 2.105 & 0.000 & 1.071 & 0.000 & 1.316 & 0.000 & 2.776 & 0.000 \\
UH-HS (direct) 
& 1.135$^{**}$ & -0.207 & 1.224 & 0.080 & 2.050 & 0.026 
& 1.014 & 0.054 & \underline{\textbf{1.008$^{**}$}} & \underline{\textbf{0.234}} & \underline{1.829} & \underline{0.341} \\
FA-AR (direct) 
& 0.876 & 0.068 & \underline{1.082$^{**}$} & \underline{0.187} & 2.106 & -0.001 
& 0.900$^{*}$ & 0.160 & 1.084$^{**}$ & 0.176 & 2.633 & 0.052 \\
FAVAR (iter) 
& 0.638$^{**}$ & 0.321 & \underline{0.877$^{*}$} & \underline{0.341} & \underline{1.955} & \underline{0.071} 
& 0.951 & 0.112 & 1.193 & 0.094 & 2.795 & -0.007 \\
DFM (iter) 
& \textbf{0.490$^{***}$} & \textbf{0.479} & \textbf{0.539$^{***}$} & \textbf{0.595} & \textbf{1.220$^{*}$} & \textbf{0.420} 
& \underline{0.549$^{**}$} & \underline{0.488} & \textbf{0.524$^{**}$} & \textbf{0.602} & \textbf{1.226$^{*}$} & \textbf{0.558} \\
BART (direct) 
& \underline{0.518$^{***}$} & \underline{0.449} & 0.966$^{*}$ & 0.274 & 1.441 & 0.315 
& \textbf{0.483$^{***}$} & \textbf{0.549} & 0.876$^{*}$ & 0.334 & \underline{1.744} & \underline{0.372} \\
\midrule
\multicolumn{13}{l}{\emph{Panel C: QWS (Tails)}}\\
AR(2) flat 
& 0.637 & 0.000 & 0.847 & 0.000 & 1.371 & 0.000 & 0.700 & 0.000 & 0.844 & 0.000 & 1.836 & 0.000 \\
UH-HS (direct) 
& 0.812$^{***}$ & -0.275 & 0.888 & -0.049 & 1.481$^{**}$ & -0.080 
& 0.717 & -0.024 & 0.757 & 0.103 & \underline{1.330} & \underline{0.275} \\
FA-AR (direct) 
& 0.619 & 0.028 & \underline{0.704$^{**}$} & \underline{0.169} & 1.375 & -0.003 
& 0.602$^{*}$ & 0.140 & \underline{0.704$^{**}$} & \underline{0.166} & 1.772 & 0.035 \\
FAVAR (iter) 
& \underline{0.455$^{**}$} & \underline{0.286} & \underline{0.678} & \underline{0.200} & \underline{1.269} & \underline{0.075} 
& 0.711 & -0.016 & 0.797 & 0.056 & 1.803 & 0.018 \\
DFM (iter) 
& \textbf{0.441$^{*}$} & \textbf{0.307} & \textbf{0.457$^{*}$} & \textbf{0.460} & \textbf{0.930} & \textbf{0.322} 
& \underline{0.519} & \underline{0.259} & \textbf{0.431$^{*}$} & \textbf{0.489} & \textbf{0.932$^{**}$} & \textbf{0.492} \\
BART (direct) 
& 0.522$^{*}$ & 0.181 & 0.898 & -0.061 & 1.253 & 0.086 
& \textbf{0.476$^{*}$} & \textbf{0.320} & 0.917 & -0.086 & 1.473 & 0.198 \\
\bottomrule
\end{tabular}}
\caption{Quantile-weighted scores by horizon and subsample (\emph{lower is better}). 
\emph{Rel.} is $1-\text{QWS}_m/\text{QWS}_{\text{AR(2)}}$ (higher is better). Superscripts denote Diebold-Mariano significance vs.\ AR(2).}
\label{tab:qws_subsamples_h612}
\end{table}

\clearpage
\section{The Inflation Driver Under Shrinkage Diagnostics from High-Dimensional Predictors }\label{sec:drivers}
With Horseshoe prior, we are able to compute $\kappa_{j}$ across MCMC draws and forecast origins and demonstrate the interpretations. For each origin we store the top-$K$ predictors by $(1-\kappa)$ and aggregate their frequency across time. This reveals persistent drivers and episodic spikes (e.g., energy, imported prices, exchange rate, labor market slack indicators). We present a heatmap of top-$K$ appearances over time (by predictor block), see \cref{sec:additionals}, and a table of overall top drivers with average $(1-\kappa)$.
For interpretation, we track the shrinkage ratio
\begin{align}
    \kappa_j = \frac{1}{1 + \tau^2 \lambda_j^2 v_j},
\end{align}
where $v_j$ is the design-scaled variance component. Small $\kappa_j$ (equivalently, large $1-\kappa_j$) signals a predictor that survives global-local shrinkage.  
We summarize $\kappa$ by origin, rank predictors by $1-\kappa$, and report the most persistent "drivers" over time. For the horseshoe regression we additionally record per-origin tables of the top-$K$, where we select $K=20$, or Top $4\%$ out of our high-dimensional predictors ranked by $(1-\kappa)$, together with cross-origin frequency summaries, to trace time-varying drivers of inflation.

\begin{figure}[ht]
    \centering
    \resizebox{\textwidth}{!}{%
        \includegraphics{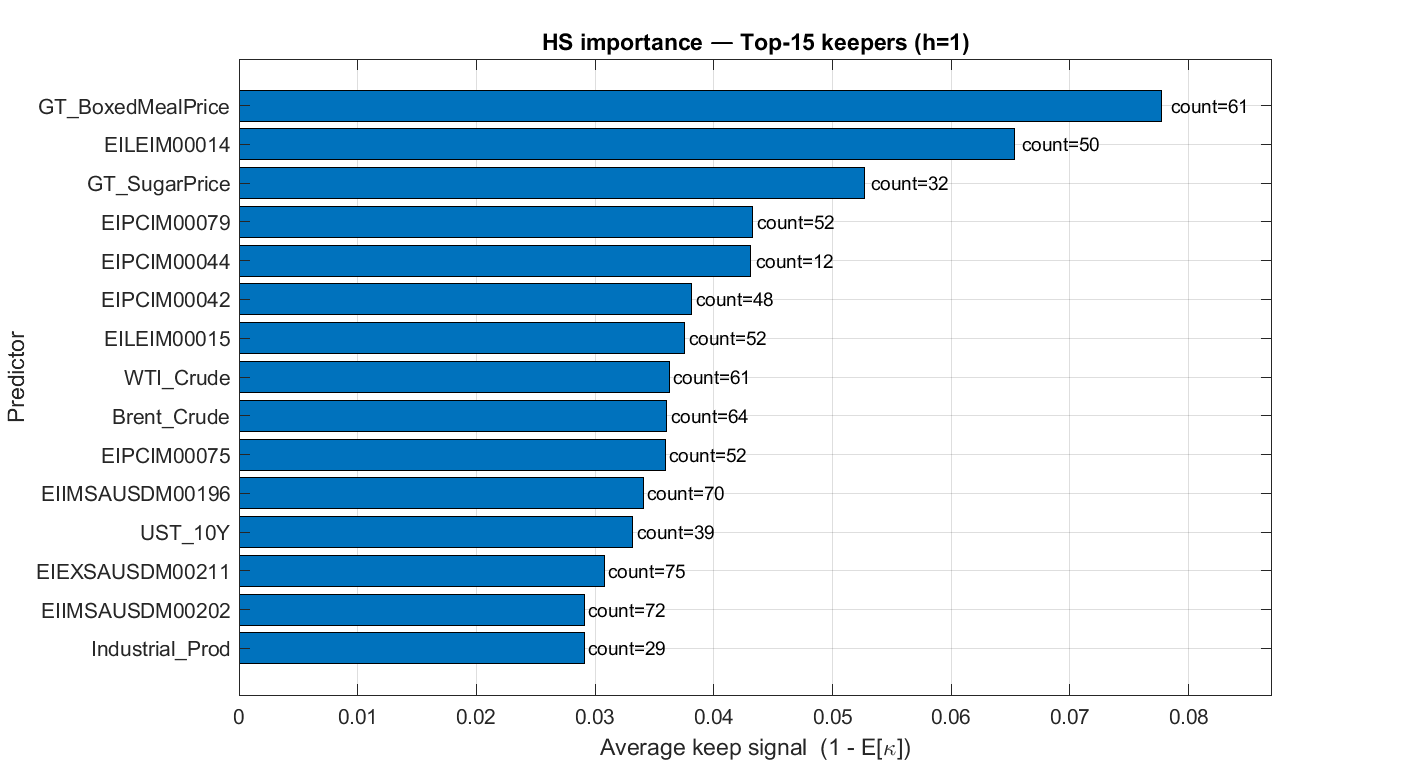}
    }
    \caption{Horseshoe average keep for forecast horizon $h=1$.}
    \label{fig:hskeepmeanh1}
\end{figure}

\begin{figure}[ht]
    \centering
    \resizebox{\textwidth}{!}{%
        \includegraphics{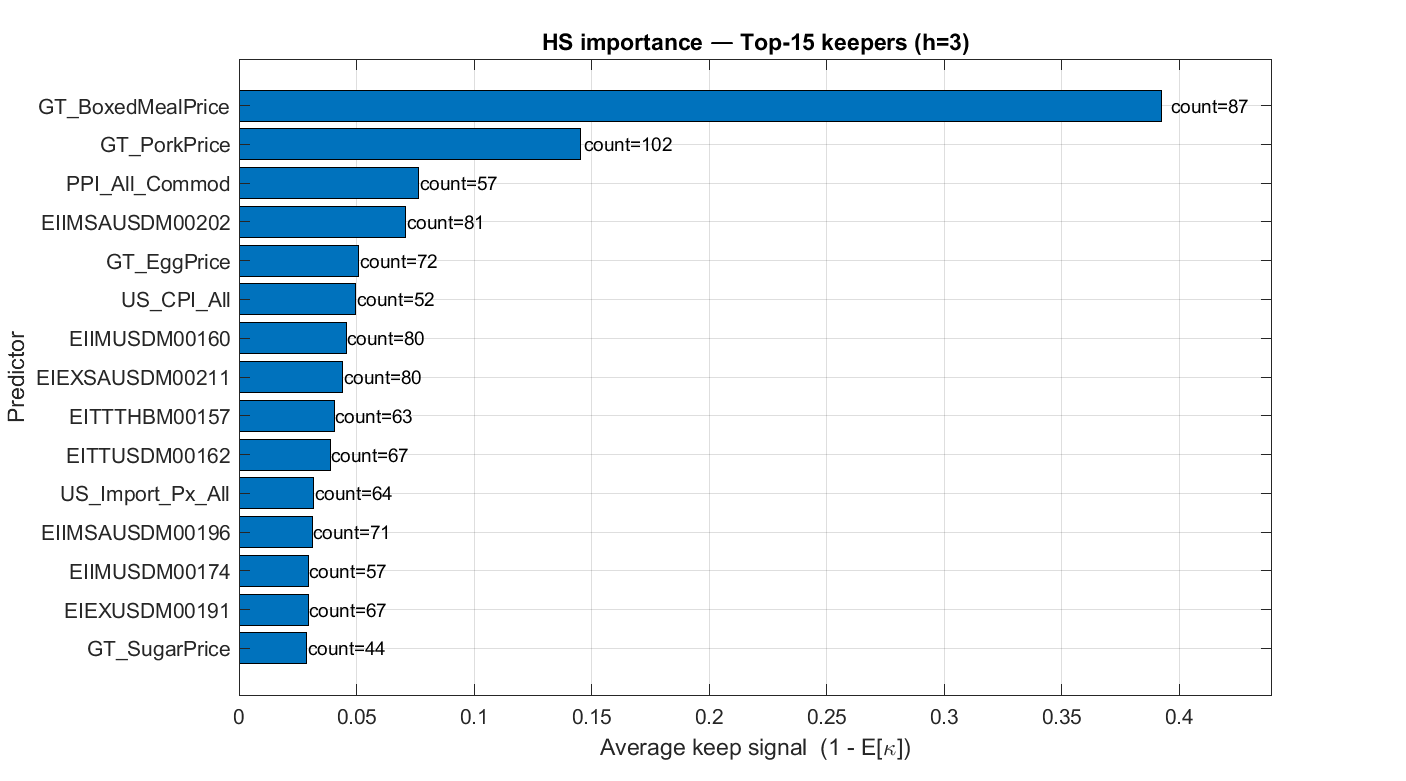}
    }
    \caption{Horseshoe average keep for forecast horizon $h=3$.}
    \label{fig:hskeepmeanh3}
\end{figure}

\begin{figure}[ht]
    \centering
    \resizebox{\textwidth}{!}{%
        \includegraphics{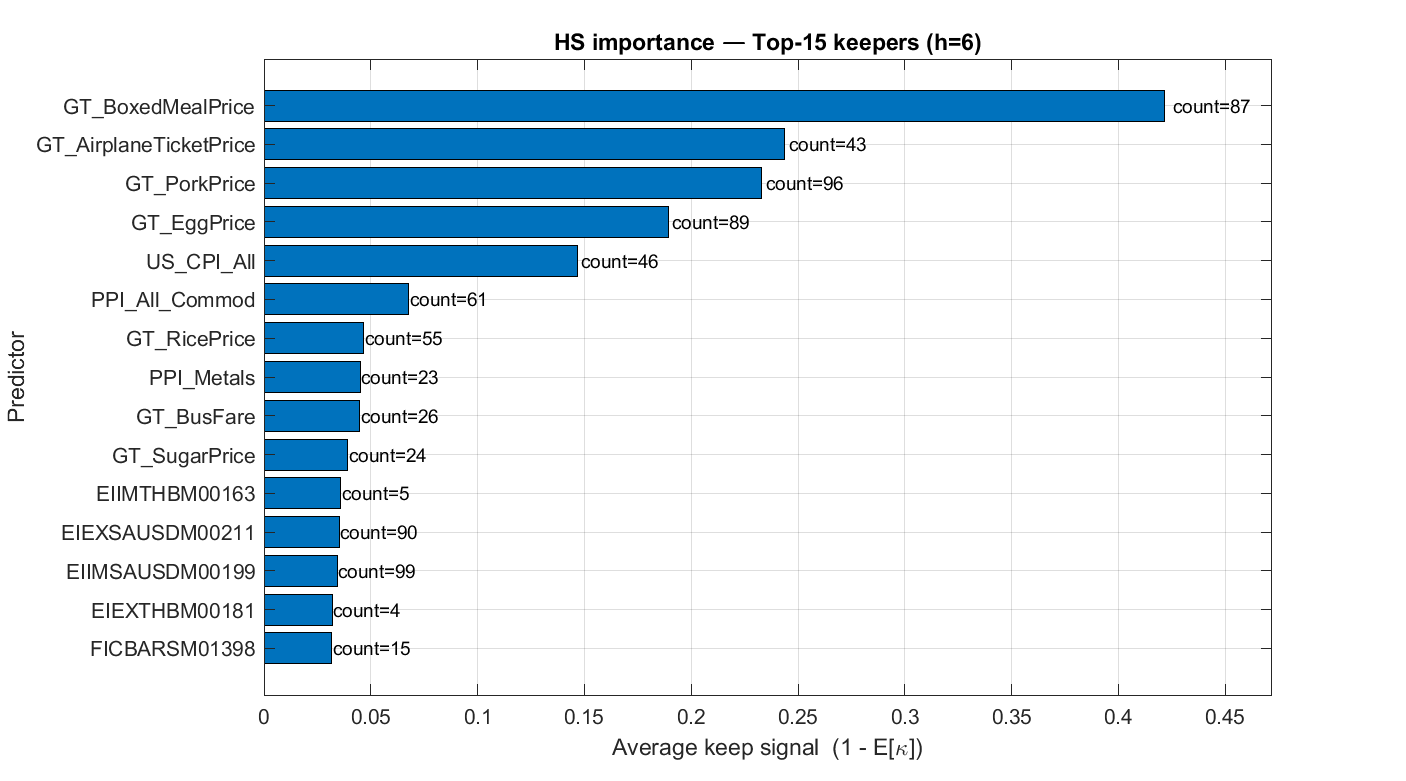}
    }
    \caption{Horseshoe average keep for forecast horizon $h=6$.}
    \label{fig:hskeepmeanh6}
\end{figure}

\begin{figure}[ht]
    \centering
    \resizebox{\textwidth}{!}{%
        \includegraphics{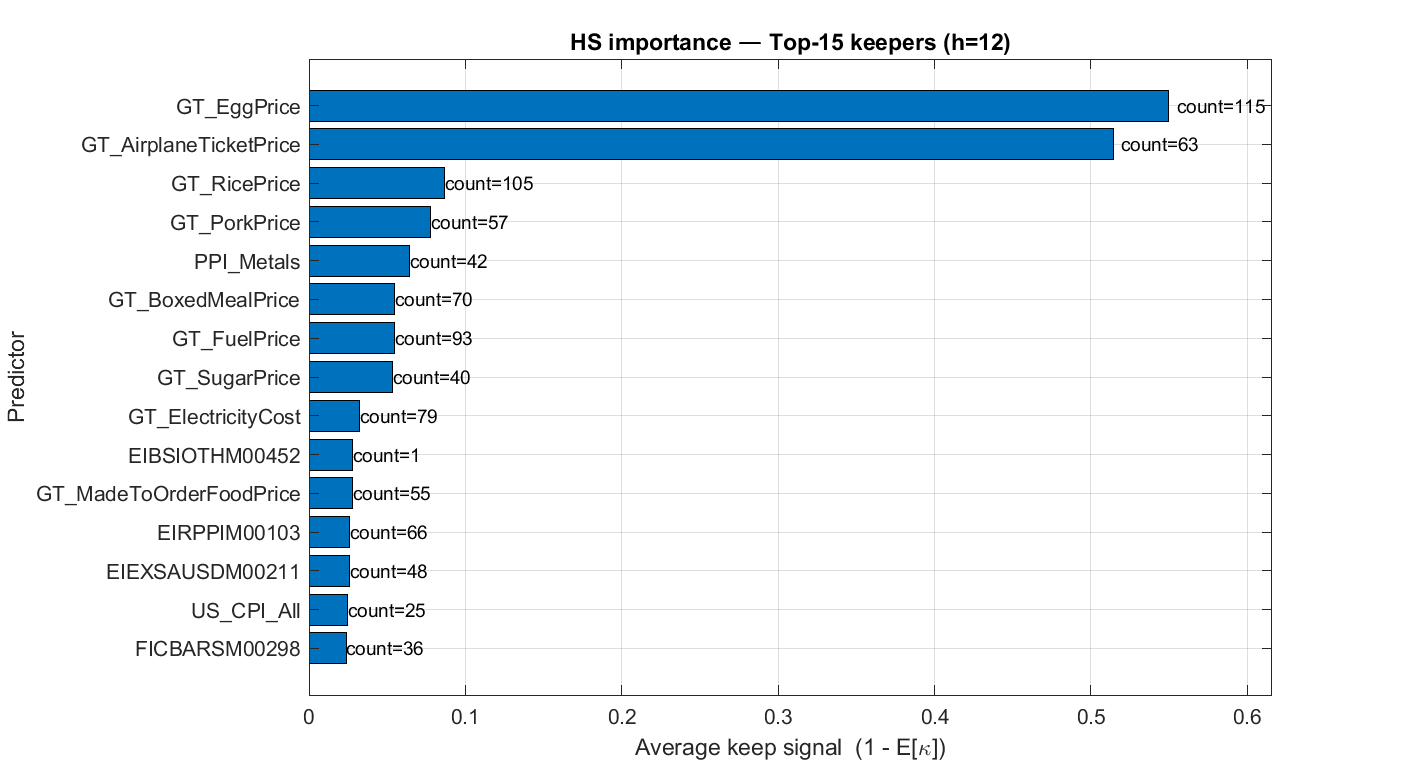}
    }
    \caption{Horseshoe average keep for forecast horizon $h=12$.}
    \label{fig:hskeepmeanh12}
\end{figure}

To open the black box of the ultra-high-dimensional horseshoe regression, we report a "keep" signal, $\mathrm{keep}_j \equiv 1-\mathbb{E}[\kappa_j\mid\text{data}]$, averaged at each forecast origin and then averaged across origins. larger bars, therefore, indicate variables that are repeatedly preserved by global-local shrinkage. Alongside the bar height, a "count" tallies how often a predictor appears in the top-20 keepers across 132 monthly forecast origins, so high counts reflect persistence rather than one-off prominence. Two caveats are important for interpretation. First, $\mathrm{keep}_j$ ranks relevance conditional on the full design and does not by itself identify structural causality. Second, with tightly collinear clusters-e.g., overlapping commodity price proxies, horseshoe typically preserves one representative at a time, so near-substitutes may rotate in and out of the top-20 even when the underlying signal is stable. Such property is essentially useful in small open economies, where multiple external indicators move together. By keeping one representative signal, the horseshoe avoids overfitting while still capturing the global cost-push component that drives domestic inflation.

\Cref{fig:hskeepmeanh1} demonstrate expected Horseshoe keep signal for the forecast horizon ($h=1$), quite obvious, the surviving predictors are dominated by the Google Trend food-away-from-home (or directly translated from Thai to English as Boxed Meal Price) and energy-cost proxies, consistent with rapid cost-push transmission into headline CPI. These categories are not unique to Thailand. They represent the typical channels through which external shocks and domestic demand interact in small open economies. Energy and import prices reflect cost-push exposure to global markets, while exchange rates and slack proxies capture transmission into local inflation \citep*{adrian2019vulnerable}. The search-intensity series for boxed-meal prices (GT\_BoxedMealPrice) tops both the average keep signal and the persistence count, and is closely followed by item-level service-sector indicators from Thailand's official data: the Services Index (seasonally adjusted (SA), series code denoted by BoT as EIPCIM00042, and EIPCIM00075, respectively) and hotel-and-restaurant activity (Sales Index SA, EIPCIM00079; VAT receipts SA, EIPCIM00044). Global energy benchmarks (WTI\_Crude, Brent\_Crude) and the Bank of Thailand's oil price inverse index (Dubai; EILEIM00015) are also repeatedly kept, highlighting the near-term role of fuel costs. Broader monetary and activity proxies-such as Broad Money (EILEIM00014) and Loans of Commercial Banks excluding interbank (FICBARSM00298)-enter with smaller average keep but nontrivial counts, indicating episodic relevance once food/energy shocks are controlled for. This composition matches institutional commentary and recent Thai experience where short-run movements in headline inflation have been driven primarily by energy and prepared-food categories, with core remaining comparatively stable \citep{BoT2025MPRQ2}.

Three months ahead ($h=3$) is illustrated in \cref{fig:hskeepmeanh3}, the model concentrates still more strongly on a narrow food-price block. GT\_BoxedMealPrice remains the single most influential driver. In this specific forecasting horizon $h=3$ we are starting to see two new predictors which contribute largely for out-of-sample prediction i.e., GT\_PorkPrice and GT\_EggPrice, which emerge as persistent survivors. GT\_SugarPrice, on the other hand, drops significantly from third to fifteenth. Pipeline cost measures-Import Price Indexes in USD, especially raw materials (EIIMUSDM00174, EIIMSAUSDM00196) and the broad PPI: All Commodities, join with US\_CPI\_All and US\_Import\_Px\_All as external price references. Export/Import Price Indexes in USD (manufactures, EIEXUSDM00191, and EIEXSAUSDM00211) and the Terms of Trade in THB/USD (EITTTHBM00157, EITTUSDM00162) appear with moderate keep and sizable counts, suggesting that imported-inflation channels are informative once the horizon extends beyond the immediate month. This pattern dovetails with evidence that a global component and foreign prices shape Thai inflation dynamics through traded inputs, while direct exchange-rate pass-through to CPI is limited and heterogeneous-findings that help explain why exchange-rate levels per se are not among the most strongly "kept" drivers once price-based import proxies are in the design \citep{manopimoke2018thai}.

At the medium horizon ($h=6$), see \cref{fig:hskeepmeanh6} for reference, transport and administered-price proxies rise in prominence. GT\_AirplaneTicketPrice becomes a persistent keeper with a high average keep signal, joined by GT\_BusFare and rice-price searches (GT\_RicePrice). Metals and all-commodity PPIs, together with import price indices in USD and THB (e.g., EIIMTHBM00163 for consumer-goods import prices in baht), continue to survive regularly, as does US\_CPI\_All. The emergence of transport-fare proxies at $h=6$ is economically intuitive for a small open economy with staggered price adjustment and policy smoothing. To aid the interpretation, energy shocks pass through gradually to administered or quasi-regulated prices for transport and utilities, which then propagate to retail and services inflation. In this window we also observe survey-based financing-cost perceptions (Other Business Sentiment such as expected interest-burden, inverted, denoted by EIBSIOTHM00452) and domestic credit (FICBARSM00298) appearing more frequently, consistent with a transition from pure cost shocks toward broader cost-of-doing-business channels as shocks mature. These horizon-specific shifts echo Thai evidence from disaggregated price data that adjustment is gradual, sectorally uneven, and more muted in core services than in fresh-food and fuel \citep*{apaitan2020thai}.

Finally at the annual horizon ($h=12$) is plotted in \cref{fig:hskeepmeanh12}, the selection stabilizes around staples and administered-price proxies with very high persistence. These are GT\_EggPrice, GT\_RicePrice, GT\_PorkPrice, GT\_AirplaneTicketPrice, GT\_FuelPrice, and GT\_ElectricityCost, which register both large keep signals and large counts. Metals PPI and broad import-price indices remain as background anchors, but the horseshoe places most weight on a compact bundle of domestic price categories that historically carry Thai CPI over year-ahead horizons. This composition is informative for the model comparison in the previous section. Precisely when the factor model's advantage fades at $h=12$, the horseshoe's focus on slow-moving staples and policy-sensitive items yields better point and density calibration, including in the tails, see \cref{tab:qws}. The prominence of import and commodity price proxies rather than the nominal exchange rate itself, aligns with Thai research showing incomplete and time-varying exchange-rate pass-through into consumer prices, i.e. the pricing-to-market and invoicing structure pushes much of the external signal into border prices and commodity indices, which our design includes directly \citep*{apaitan2024heterogeneity,Nookhwun2019FAQ141}.

Two systematic regularities cut across horizons. First, food-away-from-home and staple groceries are central at every horizon, with their relative importance rising as the horizon lengthens. This is consistent with the CPI basket's weight structure and with recent episodes in which prepared-food prices were the main contributor to core inflation movements. Second, energy and transportation proxies are most influential from $h=1$ to $h=6$ but gradually give way to staples and administered prices by $h=12$, indicating delayed pass-through and policy smoothing. The rotation among near-collinear commodity indicators (e.g., Brent vs. WTI vs. Dubai inverse) reflects horseshoe's design, in which keeping one representative from a cluster prevents overfitting without losing the underlying cost-push signal, \cite{huber2019adaptive,huber2020inducing}. These regularities square with macro evidence that Thai inflation co-moves with a global factor and supply-side developments, while domestic expectations remain relatively well anchored under inflation targeting \citep*{manopimoke2018thai}. 

Relative to prior Thai work, our contribution is twofold. Methodologically, we deliver horizon-resolved, real-time shrinkage diagnostics in an ultra-high-dimensional environment that blends official Thai series-services activity (EIPCIM00042/75), sectoral VAT (EIPCIM00044), import/export price indexes (EIIMUSDM00160, EIIMSAUSDM00196/00202; EIEXUSDM00191, EIEXSAUSDM00211), terms of trade (EITTTHBM00157/EITTUSDM00162), oil price (EILEIM00015), money and credit (EILEIM00014; FICBARSM00298), survey indicators (EIBSIOTHM00452)-with high-frequency Google Trends price proxies for staples and administered prices. Substantively, we show that the set of kept drivers is sharply horizon-dependent. To be more specific factors linked to global costs and services activity dominate near-term inflation. Additionally transport fares and broader cost burdens matter at medium horizons, and a compact group of staples and administered prices anchors year-ahead forecasts. This integrated picture helps reconcile why iterated factor models excel at $h\in\{1,3\}$ while horseshoe-regularized direct forecasts overtake them at $h=12$, and it provides policymakers with distribution-aware levers-precisely the categories that improve tail-risk calibration in our QWS results. In a literature that has emphasized global components and incomplete exchange-rate pass-through to Thai CPI, these diagnostics add transparent, data-driven evidence on which concrete price categories and official Thai series carry predictive weight at each horizon \citep*{manopimoke2018thai}.

Overall, the shrinkage diagnostics portray Thai headline inflation as a cost-push-dominated process whose drivers evolve predictably with the forecast horizon. The findings support a pragmatic forecasting strategy for Thailand, to be more specific, rely on factor-based iterated models to aggregate broad price and activity signals for the nowcast and near term, but privilege horseshoe-regularized direct specifications for medium to long horizons where a small set of staples and administered prices drive both the mean and the tails of the predictive distribution.

\section{Conclusion}\label{sec:conclude}
This paper has developed a forecasting framework for small open economies, illustrated with Thailand. Our comparison of Bayesian shrinkage and factor models demonstrates a clear horizon-dependent division of labor. Factor approaches are most effective at short horizons, when global shocks and exchange rates dominate, while shrinkage priors such as the Horseshoe become increasingly important at longer horizons. These priors stabilize inference in high-dimensional settings and deliver improved point, density, and tail forecasts.

Shrinkage diagnostics provide additional insight by revealing which predictors survive regularization. At short horizons, energy, imports, and exchange rate variables dominate. Over time, their influence recedes and domestic staples and administered prices emerge as persistent anchors. Importantly, Google Trends variables, capturing searches for food staples, rents, and daily cost-of-living items-rotate into prominence at medium and long horizons. This pattern indicates that online search behavior conveys forward-looking signals about household inflation expectations, complementing conventional macroeconomic predictors. Taken together, these dynamics underscore the need to consider both external shocks and evolving domestic sentiment when forecasting in small open economies.

Our contribution is threefold. First, we show that high-dimensional Bayesian methods are essential in environments where the number of predictors exceeds available observations and how those additional predictors play crucial roles in out-of-sample forecasts accuracy both point and density. Second, we document how the balance between global and domestic drivers evolves with the forecast horizon. Third, we provide tools-via shrinkage diagnostics-that make Bayesian forecasts interpretable for policy. Together, these results emphasize that Bayesian shrinkage and factor models are complementary, not substitutes, in the practical task of forecasting inflation in small open economies.

\section*{Acknowledgements}
The authors declare that we have no known competing financial or non-financial interests that could have influenced the research, authorship, or publication of this article. All errors are our own.

\clearpage
\appendix

\section{Additional figures and tables}
\label{sec:additionals}

\begin{figure}[ht]
    \centering
    \resizebox{\textwidth}{!}{%
        \includegraphics{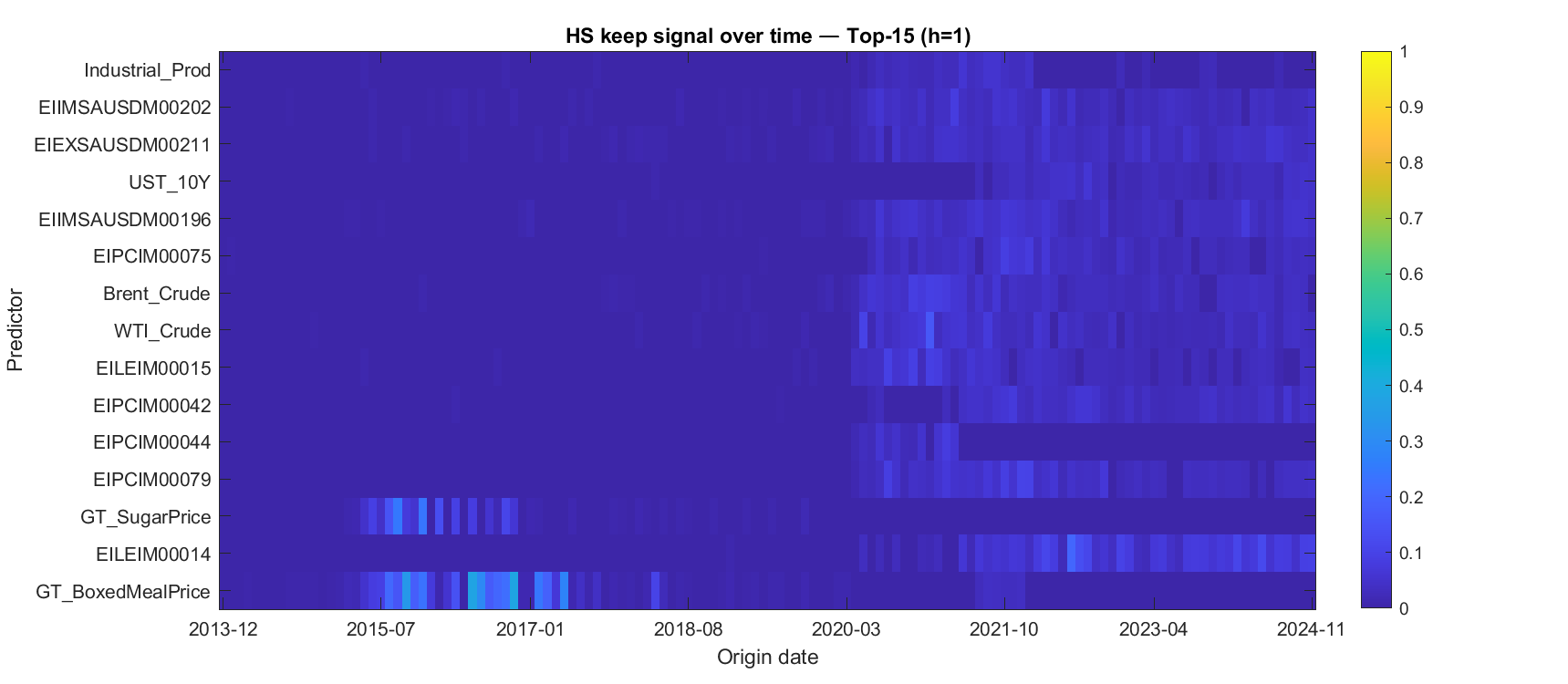}
    }
    \caption{Horseshoe average keep for forecast horizon $h=1$.}
    \label{fig:hsHeatmaph1}
\end{figure}

\begin{figure}[ht]
    \centering
    \resizebox{\textwidth}{!}{%
        \includegraphics{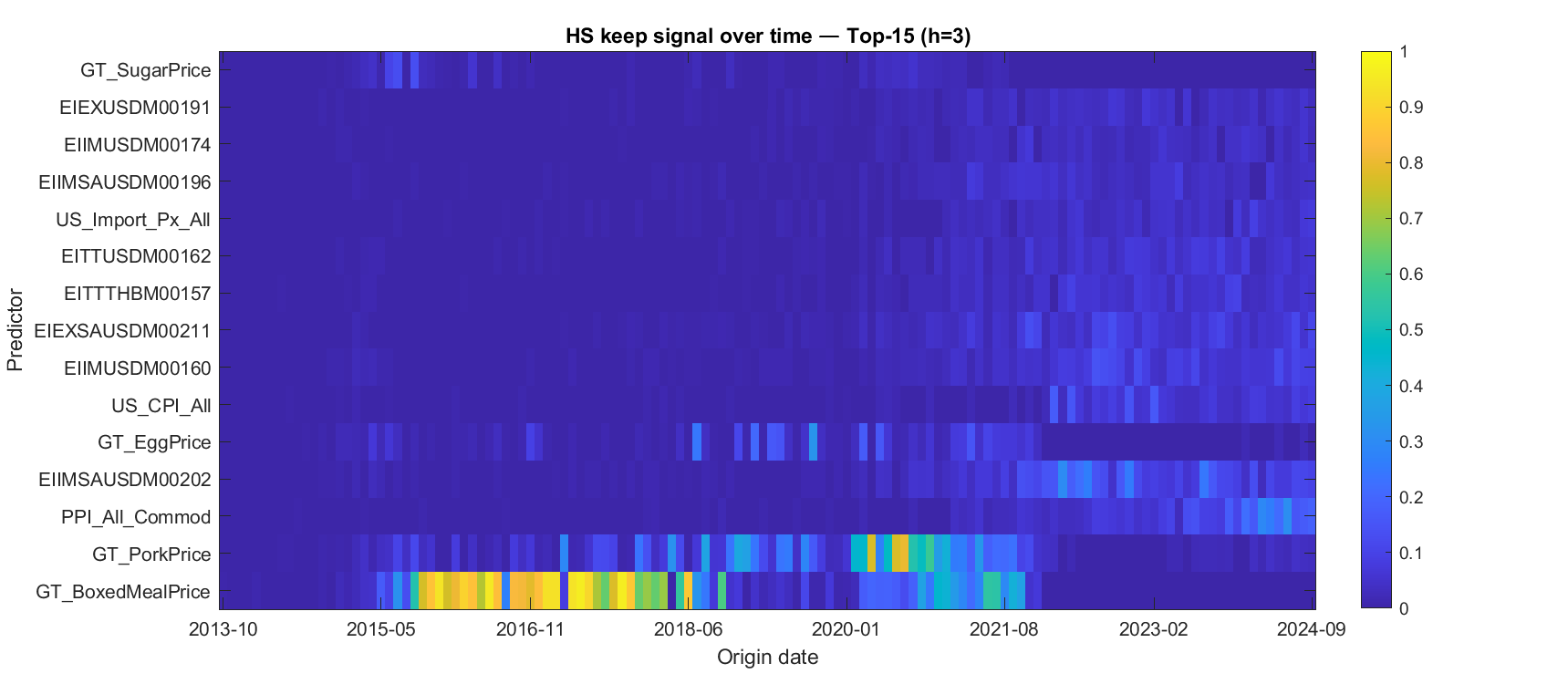}
    }
    \caption{Horseshoe average keep for forecast horizon $h=3$.}
    \label{fig:hsHeatmaph3}
\end{figure}

\begin{figure}[ht]
    \centering
    \resizebox{\textwidth}{!}{%
        \includegraphics{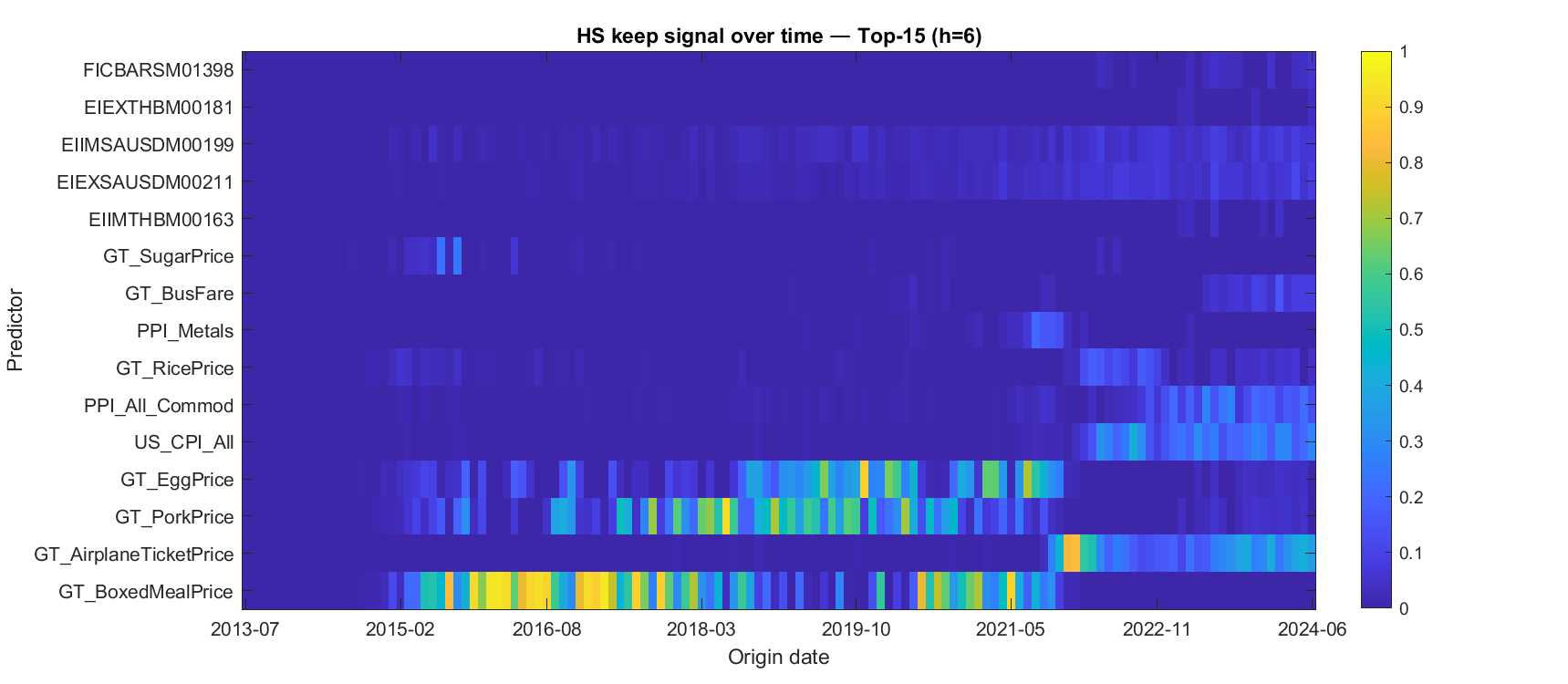}
    }
    \caption{Horseshoe average keep for forecast horizon $h=6$.}
    \label{fig:hsHeatmaph6}
\end{figure}

\begin{figure}[ht]
    \centering
    \resizebox{\textwidth}{!}{%
        \includegraphics{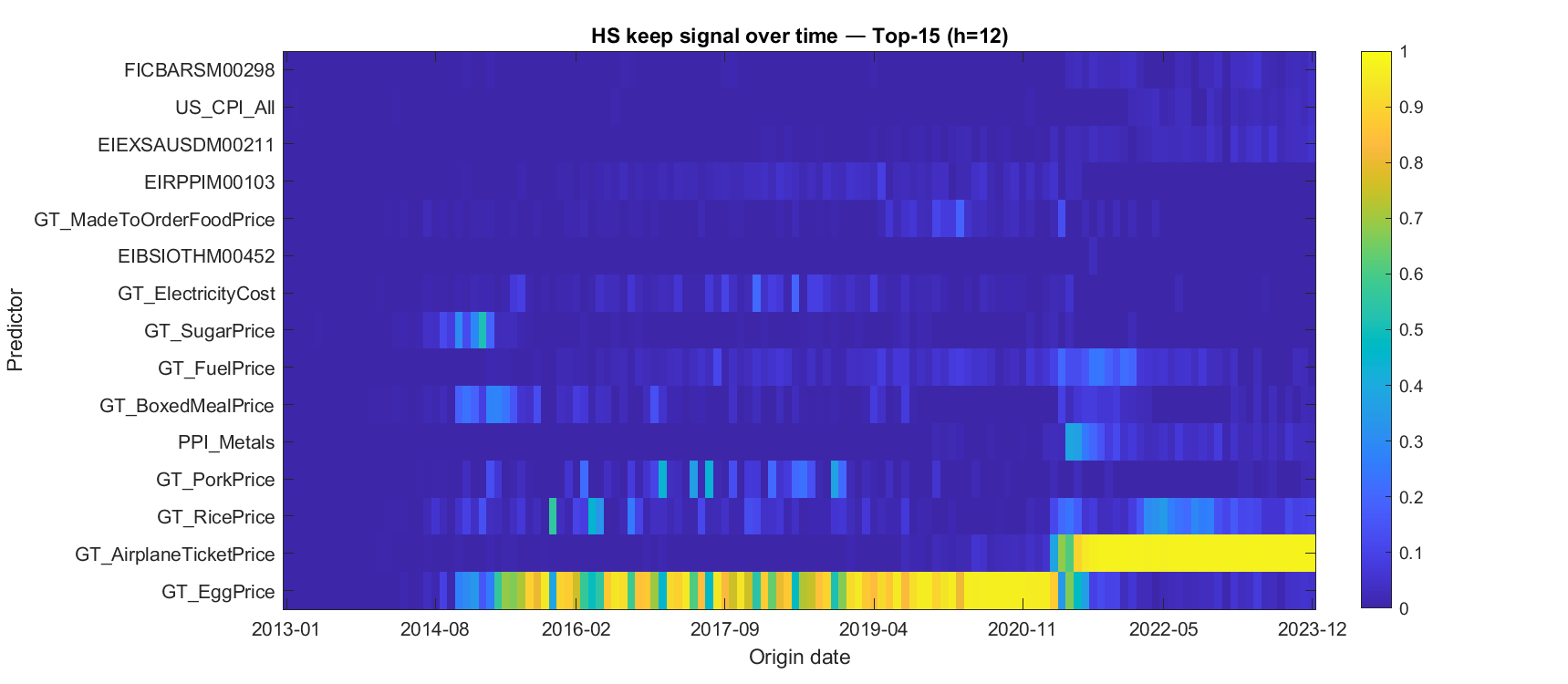}
    }
    \caption{Horseshoe average keep for forecast horizon $h=12$.}
    \label{fig:hsHeatmaph12}
\end{figure}

\clearpage
\bibliographystyle{apalike}
\bibliography{ref2}

\end{document}